\documentclass[sigconf]{acmart}

\usepackage{subfigure}
\usepackage{graphicx}
\usepackage{amsthm}
\usepackage[section]{placeins}

\newcommand{\bb}{\mathbf{b}}
\newcommand{\ba}{\mathbf{a}}

\renewcommand\footnotetextcopyrightpermission[1]{}
\newtheorem{myDef}{Definition}
\newtheorem{myTheo}{Theorem}

\AtBeginDocument{%
  }

\acmConference[]{}{}

\begin{document}

\title{FedZKP: Federated Model Ownership Verification with Zero-knowledge Proof}


\author{Wenyuan Yang}
\email{yangwy56@mail.sysu.edu.cn}
\affiliation{%
  \institution{Sun Yat-sen University}
  \city{Shenzhen}
  \country{China}
}

\author{Yuguo Yin}
\authornotemark[1]
\email{yuguoyin2002@gmail.com}
\affiliation{%
  \institution{University of Electronic Science and Technology of China}
  \city{Chengdu}
  \country{China}
}

\author{Gongxi Zhu}
\authornote{This work is done by Yuguo Yin and Gongxi Zhu during their internship at Sun Yat-sen University, under the guidance of Wenyuan Yang.}
\email{gx.zhu@foxmail.com}
\affiliation{%
  \institution{University of Electronic Science and Technology of China}
  \city{Chengdu}
  \country{China}
}

\author{Hanlin Gu}
\email{ghltsl123@gmail.com}
\affiliation{%
  \institution{WeBank}
  \city{Shenzhen}
  \country{China}
}

\author{Lixin Fan}
\authornote{Lixin Fan is the corresponding author.}
\email{lixinfan@webank.com}
\affiliation{%
  \institution{WeBank}
  \city{Shenzhen}
  \country{China}
}

\author{Xiaochun Cao}
\email{caoxiaochun@mail.sysu.edu.cn}
\affiliation{%
  \institution{Sun Yat-sen University}
  \city{Shenzhen}
  \country{China}
}

\author{Qiang Yang}
\email{qyang@cse.ust.hk}
\affiliation{%
  \institution{The Hong Kong University of Science and Technology}
  \city{Hong Kong}
  \country{China}
}

\author{}

\renewcommand{\shortauthors}{Yang et al.}

\begin{abstract}
Federated learning (FL) allows multiple parties to cooperatively learn a federated model without sharing private data with each other. The need of protecting such federated models from being plagiarized or misused, therefore, motivates us to propose a provable secure model ownership verification scheme using zero-knowledge proof (ZKP), named FedZKP. It is shown that the FedZKP scheme without disclosing credentials is guaranteed to defeat a variety of existing and potential attacks. Both theoretical analysis and empirical studies demonstrate the security of FedZKP in the sense that the probability for attackers to breach the proposed FedZKP is negligible. Moreover, extensive experimental results confirm the fidelity and robustness of our scheme.
\end{abstract}

\pagestyle{plain}



\maketitle
\section{Introduction}
Federated learning (FL) \cite{mcmahan2016federated} aims to protect data privacy by cooperatively learning a model without sharing private data among clients. 
The costly investments in training a federated model include the following aspects:  
a) it requires professional model design, high-cost dedicated hardware, and a long time to complete model training \cite{wang2021voxpopuli}; 
b) it needs a large amount of training data to improve the performance of the model to achieve better generalization ability \cite{deng2009imagenet,  chelba2013one, zhu2015aligning}. 
Therefore, verifying the ownership of valuable federated DNN models is an urgent need.

Existing schemes including FedIPR \cite{li2022fedipr} and WAFFLE \cite{tekgul2021waffle} were proposed to protect federated learning models by embedding watermarks as credentials in the models to verify the legal ownership. They are unsatisfactory in light of two drawbacks.
First, existing schemes invariably required a \textit{trusted verifier} who is responsible for checking ownership credentials sent by clients. Yet in reality, the undue expectation of trustiness in the verifier may lead to the leakage of critical credentials to attackers. The attackers can declare the ownership of models that do not belong to them, rendering these ownership verification schemes useless. 
Second, these schemes only demonstrated security against some specific attacks through empirical studies. And there is no provable security analysis from a cryptography point of view was provided. The lack of such analysis posed serious questions concerning their defense capabilities in front of potential attacks not investigated in experiments.

To address the aforementioned concerns, we propose a novel scheme called FedZKP which first embeds the hash of the exact Learning Parity with Noise (xLPN) problems \cite{jain2012commitments} into the model as the watermark and uses the solution of the xLPN problem as a credential to verify the model ownership by zero-knowledge proof (ZKP) \cite{goldwasser2019knowledge}.  To our best knowledge, this scheme is the first ZKP-based method for FL ownership verification, which immediately brings about two benefits. 
First, it allows clients to prove ownership without disclosing credentials (see Sect. \ref{Honest-Verifier Zero-Knowledge}) and, therefore, it does not require a trusted verifier to keep the credentials secret.
Second, the provable security of the FedZKP lies in the soundness of xLPN ZKP and the hardness of finding hash near-collision illustrated in Sect. \ref{Scheme Security}. As a result, the proposed FedZKP is guaranteed to defeat a variety of existing and potential attacks that are characterized by the security model in Sect. \ref{Scheme Security}. 

One crucial technical challenge addressed by the proposed FedZKP is ascribed to \textit{errors of watermarks} which may severely undermine ownership verification capabilities. These errors might result from either a normal model training process or malicious watermark removal attacks including pruning, fine-tuning and \textit{targeted destruction attack} etc. (see Sect. \ref{Validity Check} and \ref{Roboustness}).
We address the challenge by using the collision resistance hash function. Our theoretical analysis in Sect. \ref{Scheme Security} and empirical studies in Sect. \ref{Roboustness} shows that it's very hard for attackers to find a near collision and breach the defending scheme.

In summary, our contributions are threefold: 
\begin{itemize}
    \item We propose a federated model ownership verification (FMOV) scheme based on zero-knowledge proof (ZKP), which does not require a trusted verifier to protect ownership credentials (see Sect. \ref{Honest-Verifier Zero-Knowledge}). 

    \item We introduce cryptographic provable security into FMOV, by constructing a security model and giving the security proof of FedZKP under this model. Since the security model is a formal definition of attacks in FedZKP. FedZKP is proven to defeat a variety of attacks that are captured by the security model (see Sect. \ref{Sect:SecurityModel}).  

    \item Moreover, we address the technical challenge of watermark errors, using the hardness of finding hash near-collisions. Both theoretical analysis (see Sect. \ref{Scheme Security}) and experimental results (see Sect. \ref{Roboustness}) illustrate that it is difficult for the attacker to find hash near-collisions and break our scheme. 

\end{itemize}

\section{Related work}
\label{Sect:RlatedWork}
\subsection{Federated Learning}
The notion of FL was initially proposed to build a machine learning model based on datasets that are distributed across multiple parties without sharing data with each other \cite{mcmahan2016federated,mcmahan2017communication, konevcny2016federated}. The main idea is to aggregate local models learned on multiple devices, yet, without sending private data to a \textit{semi-honest} third-party server or other devices. Bearing in mind the privacy concern of secret data distributed across multiple institutions, Yang et al. extended applications of FL to a wide spectrum of use cases \cite{yang2019federated} and classified FL scenarios into three categories: \textit{horizontal federated learning}, \textit{vertical federated learning} and \textit{federated transfer learning}. Yang et al. \cite{yang2023federated} further proposed the notion of \textit{secure federated learning} to highlight the importance of  model intellectual property right when building a secure FL system.

\subsection{DNN Ownership Verification Methods}

The ownership verification of deep learning models is mainly realized through model watermarking, with a few jobs achieved through model fingerprinting \cite{10.1145/3433210.3437526,li2021modeldiff,zhao2020afa} and Proof-of-learning \cite{jia2021proof}. Model fingerprinting is a new technology aiming at extracting the model's inherent features such as specially-crafted adversarial examples to verify ownership, but its effectiveness is easily affected by changes in the model's decision boundary. Compared to Proof-of-learning, which requires storing models for different time periods and reproducing the training process, model watermarking only requires storing a model and using extraction watermark or inference to verify the ownership, and the time and space costs are lower. Therefore, we choose model watermarking to achieve ownership verification.

Existing model watermarking schemes can be divided into two categories: feature-based watermarking \cite{uchida2017embedding,fan2019rethinking,fan2021deepip,chen2018deepmarks,zhang2020passport} and backdoor-based watermarking \cite{adi2018turning,zhang2018protecting,lukas2019deep}. Considering that a backdoor attack faces a threat known as model poisoning in FL, we adopt feature-based watermarking instead of backdoor-based watermarking.

For federated learning model ownership protection, Tekgul et al. \cite{tekgul2021waffle} proposed WAFFLE, which adds a retraining step to embed the watermark after the server aggregates the local model but does not allow clients to embed and verify watermarks. Liu et al. \cite{liu2021secure} propose a backdoor-based scheme in which clients are responsible for watermark embedding under the homomorphic encryption FL framework. Another representative solution is FedIPR \cite{li2022fedipr}, which allows each client to embed their own watermarks into the model. However, the above-mentioned ownership protection schemes all use the watermark as the ownership credential, and the verification process will leak the certificate and is not provably safe.

\subsection{Zero Knowledge Proof}

Goldwasser et al. \cite{goldwasser2019knowledge} first proposed an interactive proof and zero-knowledge interactive proof system in 1985. As the research on zero-knowledge proofs intensified, zero-knowledge proofs based on LPN problems were also proposed \cite{veron1997improved}. Jain \cite{jain2012commitments} extended the LPN problem to the xLPN problem, and proposed a commitment scheme and a $\Sigma$-protocol from the xLPN problem. At present, there are also many zero-knowledge Succinct Non-interactive Arguments of Knowledge (zk-SNARK) schemes \cite{ben2013snarks,groth2016size,ames2017ligero,bunz2018bulletproofs} based on pairing, secure multi-party computing or other technologies. 

ZKP has also been applied to machine learning. Lee et al. \cite{lee2020vcnn} proposed an efficient verifiable CNN framework using ZKP, which allows clients to prove the output model is reliable without disclosing the parameters of the model. Weng et al. \cite{weng2021mystique} proposed a neural network model that supports ZKP to prove bugs in the model, check the correctness of the model output, and verify the reliability in model performance evaluation. However, there is no model watermarking schemes based on ZKP to achieve resistance to credential leakage.

\section{Preliminaries}
\label{Sect:Preliminaries}
\subsection{Notation}
We use bold lower-case and upper-case letters like $\mathbf{a, A}$ to denote vectors and matrices. $\mathbf{a}[i]$ is the $i$th bit of the binary string $\mathbf{a}$. The Hamming weight of $\mathbf{a}\in\{0,1\}^{m}$ is denoted as $||\mathbf{a}||_1=\sum^{m}_{i=1}\mathbf{a}[i]$. $\{0,1\}^m$ means the set of all binary strings of length $m$. $\{0,1\}^m_a$ means the set of all binary strings of length $m$ and Hamming weight $a$. The Hamming distance between $\mathbf{a,b}\in\{0,1\}^m$ is computed as $\sum^{m-1}_{i=0}(\mathbf{a}_i\oplus\mathbf{b}_i)$. The concatenation of vectors $\mathbf{a}$ and $\mathbf{b}$ is written as $\mathbf{a||b}$. $a\stackrel{R}{\leftarrow}A$ denotes $a$ is selected uniformly random from set $A$. Define a permutation $\pi$ on the vector $\ba\in\{0,1\}^{m}$ as $\bb \triangleq \pi(\mathbf{a})$ such that $\mathbf{b}[\pi(i)] = \mathbf{a}[i]$ for any $i\in\{1, \cdots, m\}$. Probabilistic polynomial time (PPT) algorithms are written like the hash function $\mathtt{H}(\cdot)$. $\lfloor a\rceil$ means rounding to the nearest whole number for decimal $a$, and $\lfloor a\rfloor$ means rounding down for decimial $a$. The main notations used in this paper are shown in Tab. \ref{Main Notations Used in This Paper}.

\begin{table}[!ht]
    \centering
    \caption{Main Notations Used in This Paper}
    \begin{tabular}{|c|c|}
    \hline
    Notations & Descriptions\\
    \hline
    ${K}$ & Number of clients in Federated Learning\\
    \hline
    $\mathbb{M}$ & Federated Neural network model\\
    \hline
    $\mathbf{A}$ & The xLPN matrix\\
    \hline
    $\mathbf{s}$ & The xLPN secret vector (Ownership credential)\\
    \hline
    $\mathbf{e}$ & The xLPN error vector\\
    \hline
    $\mathbf{y}$ & The xLPN public vector\\
    \hline
    $m$ & The row number of xLPN matrix\\
    \hline
    $l$ & The column number of xLPN matrix\\
    \hline
    $\mathcal{B}$ & Bernoulli distribution\\
    \hline
    $\tau$ & The parameter of Bernoulli distribution\\
    \hline
    $\mathbf{A}_{agg}$ & The aggregated xLPN matrix\\
    \hline
    $\mathbf{y}_{agg}$ & The aggregated xLPN public vector\\
    \hline
    $\mathtt{H}(\cdot)$ & Hash function\\
    \hline
    $\mathtt{Com}(\cdot)$ & Function to generate commitment\\
    \hline
    $\mathbf{h}$ & Hash watermark\\
    \hline
    ${n}$ & Output length of hash function\\
    \hline
    $err_n$ &  Security Boundary of Watermark Detection\\
    \hline
    $\pi$ & The random permutation\\
    \hline
    $\leftarrow$ & Sampling from a set\\
    \hline
    $\mathbf{E}$ & Watermark embedding matrix\\
    \hline
    $\mathcal{I}$ & System initialization process\\
    \hline
    $\mathcal{K}$ & Watermark generation process\\
    \hline
    $\mathcal {V}$ & Ownership verification process\\
    \hline
    $\mathcal{W}$ & Watermark embedding process\\
    \hline
    $\mathbf{D}$ & Dataset of model training\\
    \hline
    $\mathcal{L}$ & Watermark location parameter\\
    \hline
    $\mathbf {W}_{\gamma}$ & Weights of normalization layer\\
    \hline
    \end{tabular}
    \label{Main Notations Used in This Paper}
\end{table}

\subsection{Horizontal Federated Learning}
The federated learning system composed of $K$ clients and the server is implemented as the following steps:
\begin{itemize}
    \item Each client $c_i, i\in\{1,2,...,K\}$  uses its own datasets $\mathcal D_i$ to update new model  parameters $\mathbf W_{k}$ by optimizing the model main task and send $\mathbf W_{k}$ to the server.
    \item The server aggregates the received local model parameters to  average global parameters $\mathbf W$  as the following Eq. (\eqref{FLavg}) and distributes them to each client. 
    \begin{equation}
   \mathbf W = \sum\limits_{k=1}\limits^K \frac{\lambda_k}{K}\mathbf W_{k}, 
   \label{FLavg}
\end{equation}
where $\lambda_k$ is the weight of $\mathbf W_{k}$ and $\sum\limits_{k=1}\limits^K  \lambda_k =K$. 

\begin{figure}
    \centering
    \includegraphics[scale=0.23]{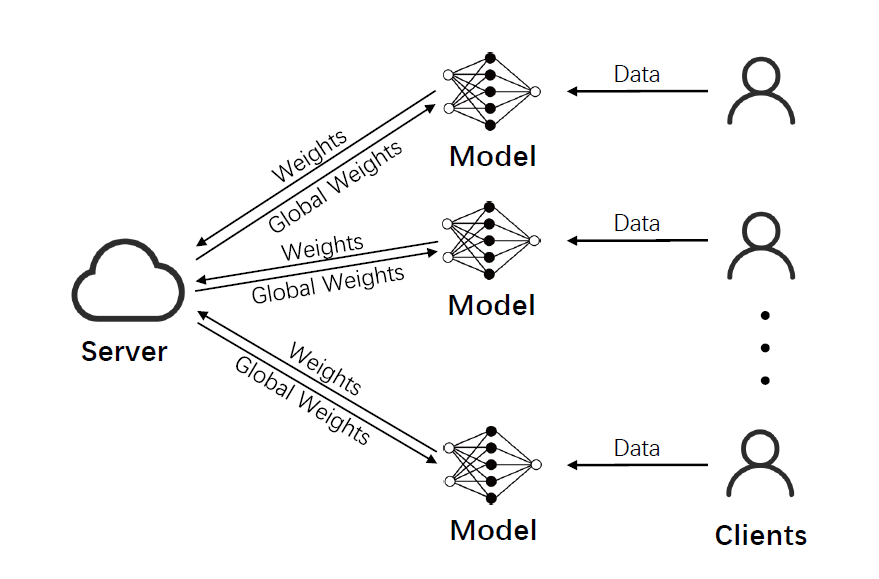}
    \caption{Federated Learning System}
    \label{Federated Learning System}
\end{figure}
\end{itemize}
These two steps iterate until the performance of the federated model does not improve.

\subsection{Learning Parity with Noise}
\begin{myDef}[\textbf{Search LPN Problem}]
    \label{LPN Problem}
    For $m,l\in \mathbb{N}$, the search $(m, l)$-LPN$_\tau$ problem is $\epsilon$-hard if for every PPT algorithm $\mathcal{A}$:
\end{myDef}

\begin{equation}
    \bigg|\text{Pr}[\mathcal{A}(\mathbf{A}, \mathbf y = \mathbf{A\cdot s}\oplus \mathbf{e})=\mathbf{s}]\bigg|\leq\epsilon
\end{equation}

\textsl{where $\mathbf{A}\stackrel{R}{\leftarrow}\{0,1\}^{m\times l}, \mathbf{e}\stackrel{R}{\leftarrow}\mathcal{B}_{\tau}^{m}$, and $\mathbf{s}\stackrel{R}{\leftarrow}\{0,1\}^{l}$. $\mathbf{A}$ is the LPN matrix. $\mathbf{y}$ is the LPN public vector. $\mathbf{e}$ is the LPN error vector selected randomly from Bernoulli distribution $\mathcal{B}_{\tau}^{m}$. $\tau$ is the parameter of the Bernoulli distribution. $\mathbf{s}$ is the LPN secret vector.}

In the search LPN problem, the error distribution is the Bernoulli distribution $\mathcal{B}_{\tau}^{m}$ with a parameter $0\textless\tau\textless\frac{1}{2}$. Specifically, for $1\textless i\textless m$, the $\mathbf{e}_i$ is chosen independently and distributed identically with Pr[$\mathbf{e}_i=1$]=$\tau$. The search LPN$_\tau$ is the same as the problem of decoding random linear codes, which is believed to be exponentially hard.

\begin{myDef}[\textbf{Search Exact LPN Problem}]
    \label{Exact LPN Problem}
    The search exact LPN (xLPN) problem is defined by Jain \cite{jain2012commitments}. The search xLPN problem is similar to the search LPN problem, except that the Hamming weight of the error vector $\mathbf{e}$ is exact $\lfloor m\tau \rceil$, which means $\mathbf{e}$ is sampled uniformly at random from the set $\{0,1\}^{m}_{\lfloor m\tau \rceil}$.
\end{myDef}

If the search xLPN$_\tau$ problem is easy, the search LPN$_\tau$ problem will be solved, the difficulty of the search xLPN$_\tau$ problem is not easier than the difficulty of the search LPN$_\tau$ problem.

The \textbf{public input} of search xLPN problem is $\mathbf{A}$ and $\mathbf{y}$. $\mathbf{A}$ and $\mathbf{y}$ are the xLPN problem instance to be solved in verification, and their hash value will be used as the watermark of the model. The \textbf{private input} of xLPN problem is $\mathbf{s},\mathbf{e}$, which is the credential in FedZKP.

\subsection{Commitment Scheme}
The commitment scheme $\mathtt{Com}$ is a two-party interactive protocol between a sender and a receiver. First, the key generation algorithm outputs a public commitment key $pk$. Next, the sender computes a commitment value $(c,d)=\mathtt{Com}(m, pk)$ and sends $c$ to the receiver, where $m$ is a string and $d$ is the random opening. Finally, the sender sends a random opening $d$ and a message $m$ to the receiver, and the receiver verifies $(c,d)=\mathtt{Com}(m, pk)$ with the public key. If it satisfies, the receiver outputs 1, and the receiver believes that $c$ is a commitment to $m$ under the public key $pk$. Else it outputs 0, and the receiver believes that $c$ is not a commitment to $m$ under the public key $pk$.
The commitment scheme satisfies the following two properties:
\begin{itemize}
    \item $\textbf{Perfect Binding}$: It's infeasible for every commitment $c$ to be opened to two different messages.
    
    \item $\textbf{Computational Hiding}$: The commitment $c$ hides the message computationally. For every different message $m,m'$ with equal length and $(c,d)\stackrel{R}{\leftarrow}\mathtt{Com}(m,pk)$, $(c',d')\stackrel{R}{\leftarrow}\mathtt{Com}(m',pk)$, the distribution of $c$ and $c'$ are computationally indistinguishable.
\end{itemize}

\subsection{$\Sigma$-Protocol}
$\Sigma$-Protocol \cite{damgaard2002sigma} is a kind of method to design ZKP. Let ($\mathtt P$, $\mathtt V$) be a two-party protocol, where $\mathtt P$ and $\mathtt V$ are Probabilistic polynomial
tim (PPT) algorithms, and let $\mathcal{R}$ be a binary relation. Then the ($\mathtt P$, $\mathtt V$) is called a $\Sigma$-Protocol for relation $\mathcal{R}$ under challenge set $C$, public input $y$ and private input $w$, if and only if it satisfies the following conditions:
\begin{itemize}
    \item $\textbf{3-Move Form}$: The protocol is of the following form:
    \begin{itemize}
        \item $\mathtt P$ computes a commitment $t$ and sends it to V.
        \item $\mathtt V$ selects a random number $c\stackrel{R}{\leftarrow}C$ and sends it to $\mathtt P$.
        \item $\mathtt P$ sends the response $s$ to the verifier.
        \item $\mathtt V$ accepts or rejects the proof depending on the protocol transcript $(t,c,s)$.
    \end{itemize}
    \item $\textbf{Completeness}$: The V always accepts if $(y,w)\in\mathcal{R}$.
    \item $\textbf{Special Soundness}$: There is a PPT knowledge extractor $\mathcal{E}$ which takes a set $\{(t,c,s)|c\in C\}$ of transcripts with the same commitment as inputs, and outputs $w'$ so that $(y, w')\in\mathcal{R}$.
    \item $\mathbf{Special\ Honest-Verifier\ Zero-Knowledge}$: There are PPT simulator $\mathcal S$ taking $y$ and $c\in C$ as inputs, and outputs a triple $(t,c,s)$ whose distribution is computationally indistinguishable from accepting protocol transcripts generated by a real protocol runs.
\end{itemize}

\subsection{Near Collision}
Collision resistance is one of the guidelines of the provably secure hash function \cite{menezes2001handbook}. The function $\mathtt{Dif}(\mathbf{a},\mathbf{b})=\sum^{n}_{i=1}(\mathbf{a}[i]\oplus\mathbf{b}[i])$ denotes the number of different bits between two equal-length binary vectors $\mathbf{a,b}$.

Let $\mathcal{H}=\{\mathtt{H}:\{0,1\}^{*}\rightarrow\{0,1\}^{n}\}$ be a family of hash functions. For $n'\in\{1,...,n\}$ and $\mathbf{x}\in\{0,1\}^{*}$, we say that $n'$-near collision happens between $\mathbf{x}$ and $\mathbf{x}'\in\{0,1\}^{*}$ if the Hamming distance between $\mathtt{H}(\mathbf{x})$ and $\mathtt{H}(\mathbf{x}')$ is less than $n'$.
\section{Problem Definition}
\label{Sect:Problem Definition}
\subsection{System Model}
The system model focuses on the role of each entity in FMOV. In FedZKP, there are three kinds of entities i.e., client, server, and verifier.
\begin{itemize}
    \item \textbf{Clients} build their own local models using their own datasets, generates their own xLPN public inputs from credentials, and embed the watermark into the local models.
    \item The \textbf{server} generates the watermark from clients' xLPN public inputs and aggregates the clients' local models into a global model.
    \item The \textbf{verifier} performs ownership verification on the models suspected of being stolen, whose identity changes with realistic scenarios, including courts, notary public, or a user who wants to verify ownership before using another's model.
\end{itemize}

Moreover, the entity that wants to prove ownership of the model in the verification is the prover, which may be a client or an attacker.

In this scheme, the clients contribute valuable datasets and computing resources. They should generate credentials and claim ownership. The server just coordinates training and aggregation, which doesn't make enough contributions to claim ownership.

\subsection{Threat Model}
\label{Threat Model}
The threat model focuses on the trust assumptions about the entities in FMOV and the capabilities of the attacker. We assume that the federated learning process is executed and completed normally as set. We make the following trust assumptions about the entities in the FMOV:
\begin{itemize}
    \item \textbf{Clients} are trusted and perform the steps in the scheme correctly to protect their ownership benefits.
    \item \textbf{Server} is honest but curious. The server performs watermark generation and watermark embedding correctly, but tries to steal the credentials.
    \item \textbf{Verifier} is honest but curious.The verifier will be honest in publishing ownership verification results, but tries to get credentials during verification.
\end{itemize}

The attacker in this scheme could be the entity inside or outside the FL system including the server, verifier, and other entities outside the FL system. The attacker aims to steal models with high accuracy, and uses his capabilities to pass ownership verification on a model that does not belong to him. The capabilities of the attacker in FMOV are listed as follows:

\begin{itemize}
    \item Since the ownership verification is public, the attacker can obtain information in ownership verification, including model watermarks, public inputs, parameters transmitted in ZKP, etc.
    \item The attacker steals some models and obtains their structures, weights, and watermarks. He can modify the watermarks of the stolen models by fine-tuning, pruning, and targeted destruction attack, trying to pass ownership verification with private inputs generated by the attacker, and interfering with the ownership proof of the real client.
\end{itemize}



\section{Proposed Scheme}\label{Sect:ProposedScheme}
\begin{figure*}
    \centering
    \includegraphics[scale=0.45]{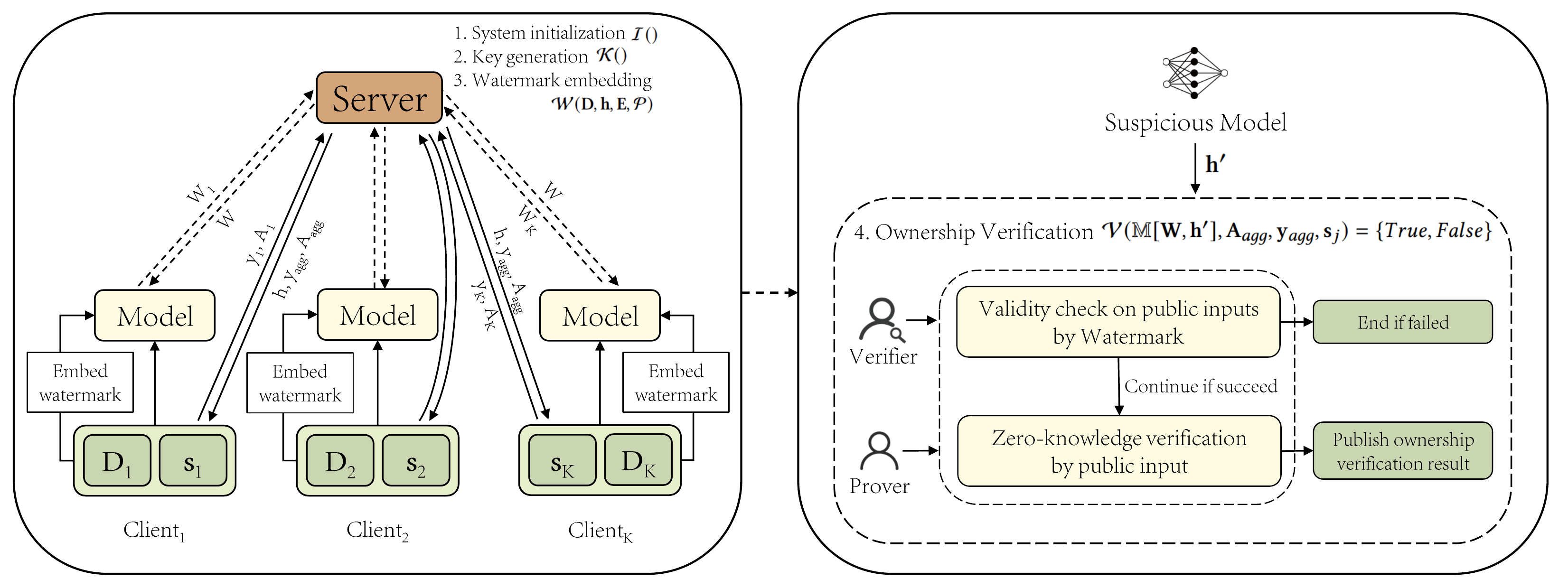}
    \caption{FedZKP consists of four processes:  \textit{System Initialization $\mathcal I$} , \textit{Watermark Generation $\mathcal K$}, \textit{Watermark Embedding $\mathcal W$}, and \textit{Ownership Verification $\mathcal V$}. Client $c_i$ has dataset $\mathbf D_i$ and credential $\mathbf{s}_i$. Client $c_i$ sends the public inputs $\mathbf{A}_i,\mathbf{y}_i$ to the server, and gets watermark $\mathbf{h}$, aggregated public inputs $\mathbf{A}_{agg}$ and $\mathbf{y}_{agg}$ from the server. $\mathbf{W}_i$ is the local model of client $c_i$. $\mathbf{W}$ is the global model.}
    \label{Fig:overview}
\end{figure*}

The overview of FedZKP is shown in Fig. \ref{Fig:overview}. In FedZKP, the hash value of the public inputs in ZKP is embedded into the model as the watermark,
and the private input in ZKP is used as a credential to verify
the ownership of the model.
In this section, we will describe our FedZKP scheme which includes four steps as Def. \ref{FedZKP Definition}: \textit{System Initialization ($\mathcal I$)}, \textit{Watermark Generation ($\mathcal K$)}, \textit{Watermark Embedding ($\mathcal W$)}, and \textit{Ownership Verification ($\mathcal V$)}. In Sect. \ref{System Initialization}, public parameters of the system are initialized. in Sect. \ref{watermark_generation}, the clients generate the xLPN public inputs from credentials, and send the public inputs to the server to get the watermark. In Sect. \ref{watermark_embedding}, clients embed the watermark into the federated model following FedIPR \cite{li2022fedipr}. In Sect. \ref{ownership_verification}, the client uses the credential to prove the ownership of the federated model to the verifier. Because of the zero-knowledge property of ZKP, our scheme does not require a trusted verifier to ensure the confidentiality of credentials.

\begin{myDef}
    \label{FedZKP Definition}
    A federated ownership verification scheme with zero-knowledge proof (FedZKP) is a tuple $\mathcal T=(\mathcal{I}, \mathcal K, \mathcal W,\mathcal V)$ of the process:
\end{myDef}

\begin{figure}
    \centering
    \includegraphics[scale=0.20]{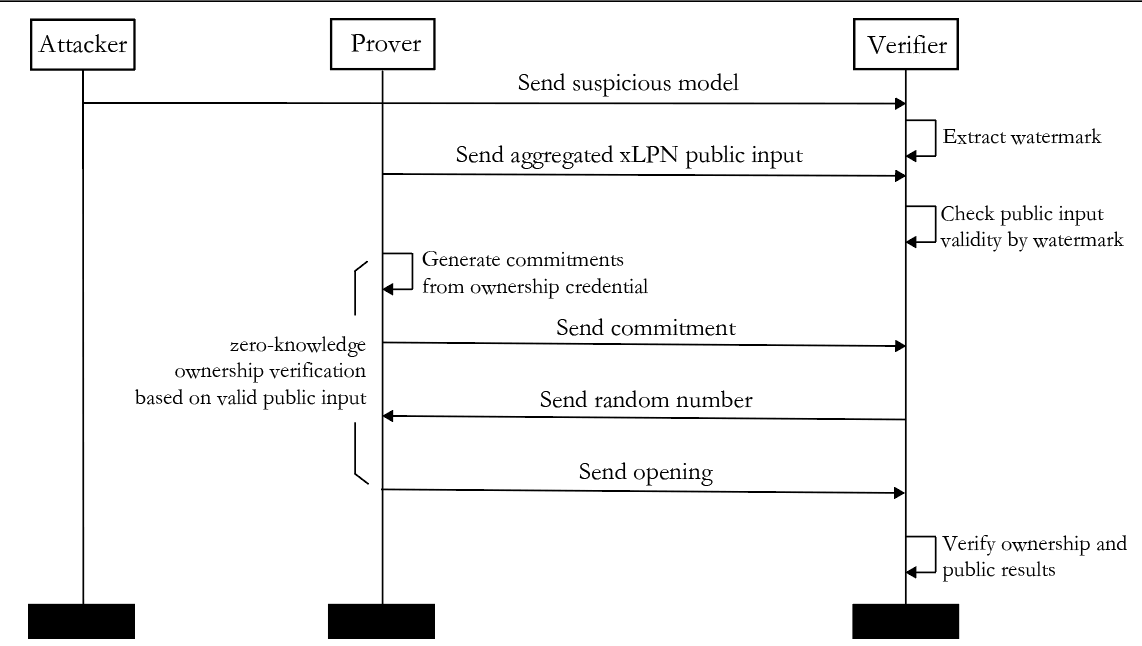}
    \caption{Sequence of FedZKP Ownership Verification}
    \label{Fig:verification sequence diagram}
\end{figure}

\begin{itemize}
    \item [1.] System initialization process $\mathcal{I}() \rightarrow PP $ sets the initial public parameters of the system before the federated system operates, in which $PP$ is prepared public parameter set.

    \item [2.] Watermark generation process $\mathcal K() \rightarrow (\mathbf s,\mathbf{h})$: For a federated learning system with $K$ clients, each client $ j \in \{1,\dots,K\}$ generates its own ownership credentials $\mathbf{s}_j$, the public matrix $\mathbf{A}_j$, and public parameter $\mathbf y_j$. The server uses $(\mathbf{y}_1, ... , \mathbf{y}_K)$ and $(\mathbf{A}_1, ..., \mathbf{A}_K)$ to generate a hash watermark $\mathbf{h}$.
    
    \textbf{Remark:} $\mathbf s=(\mathbf{s}_1,...,\mathbf{s}_K)$ is the private credentials held by each client, which needs to be kept secret for the server and other clients. For subsequent zero-knowledge ownership verification, $\mathbf{h}$ needs to be made public.

    \item [3.] Watermark embedding process $\mathcal W( \mathbf D,\mathbf{h}, \mathbf{E}, \mathcal P)=\mathbb M[\mathbf{W},\mathbf{h}]$ is accompanied by model training, and the final model $\mathbb M[\mathbf W,\mathbf{h}']$ is obtained by minimizing the combined loss $\mathtt {L}$ (defined in Eq. \eqref{combined_loss} ) in which the main task $\mathtt {L}_D$ is followed by a regularization term $\mathtt {L}_\mathbf{h}$ aimed at embedding watermarks.
    
    \begin{equation}
        \label{combined_loss}
        \mathtt L=\mathtt{L}_D(\mathtt f(\mathbf W,\mathbf {X}_d),\mathbf {Y}_d)+\lambda \mathtt {L}_{\mathbf{h}}(\mathbf W_\gamma, \mathbf{h})
    \end{equation}
    
    where $\mathbf {X}_d$ represents the data used by the model training, $\mathbf{Y}_d$ is the label of $\mathbf {X}_d$ , $\mathtt f() $ is the model inference result, $\mathbf {W}_\gamma $ is the target parameter for embedding the watermark, and $\lambda$ is the weight parameter of $\mathtt {L}_\mathbf{h}$.


    \item [4.] The Ownership verification $\mathcal V(\mathbb {M}[\mathbf W,\mathbf {h'}]$, $ \mathbf{A}_{agg}, \mathbf{y}_{agg},\mathbf{s}_j)=\{True, False\}$ allows the client $j$ to use the credential $\mathbf{s}_j$ to prove whether he has ownership of the given model $\mathbb M[\mathbf{W},\mathbf{h'}]$ through the zero-knowledge proof protocol. The sequence diagram of FedZKP ownership verification is shown in Fig. \ref{Fig:verification sequence diagram}.
\end{itemize}

\subsection{System Initialization}
\label{System Initialization}
This step is used to set the initial public parameters $PP$ of the system. The $PP$ is a standard of the scheme, which can be either a parameter published by the Standards Institute or a negotiation between federations. The elements of public parameter $PP$ in FMOV contains

$$\bigg(\mathcal{P}, \mathbf{E}, m, l, 0\leq\tau\leq\frac{1}{2}, w, p_r, err_{n}\bigg).$$

\begin{itemize}
    \item $\mathcal{P}$ is a positional parameter that determines which layers of the model the watermark is embedded in.
    \item $\mathbf{E}$ is the embedding matrix and $\mathbf E^{\omega \times n} \stackrel{R}{\leftarrow}\mathcal{N}(0,1)$ in which $\omega$ is the number of model parameters which watermark is embedded in, $n$ is the length of the hash watermark and $\mathcal{N}(0,1)$ is a standard normal distribution.
    \item $m$ is the row number of the xLPN public matrix while $l$ is the column number. $\tau$ is the parameter of the Bernoulli distribution $\mathcal{B}^{m}_\tau$ in xLPN.
    \item $w={\lfloor m\tau \rceil}$ is the Hamming weight of the xLPN error vector. 
    \item $p_r$ is the required probability of finding near-collision for a single computation of the hash value in the system.
    \item $err_n$ is the security boundary of the watermark detection errors (see Def. \ref{Def:Security Boundary}). It denotes the upper limit of the number of hash watermark bits an attacker can damage when $p_r$ and the hash function output length $n$ are given.
\end{itemize}

\subsection{Watermark Generation}
\label{watermark_generation}
One of our innovations different from the previous federated ownership protection scheme \cite{li2022fedipr, tekgul2021waffle, liu2021secure} is that FedZKP does not use a watermark as an ownership credential, but uses the private input  generated by each client as the ownership credential of each client. 
Specifically, watermark generation process $\mathcal K() \rightarrow (s,\mathbf{h})$ can be divided into the following steps:
\begin{itemize}
    \item [1. ] For a federated learning system, each client $j\in \{ 1, \dots, K\}$ generate a public matrix  $\mathbf{A}_j\stackrel{R}{\leftarrow}\{0,1\}^{m\times l}$, a secret vector $\mathbf{s}_j\stackrel{R}{\leftarrow}\{0,1\}^{l}$ and a noise vector $\mathbf{e}_j\stackrel{R}{\leftarrow}\{0,1\}^{m}_{\lfloor m\tau \rceil}$. It computes an xLPN public input: 

    \begin{equation}
        \mathbf{y}_j=\mathbf{A}_j\cdot\mathbf{s}_j\oplus \mathbf{e}_j.
        \label{generate_y_j}
    \end{equation}

    Each client sends its public matrix and xLPN public input pair $(\mathbf{A}_j,\mathbf{y}_j)$ to the server. 

    \item[2. ] After receiving all the pairs sent by every client, the server uses these pairs to form the aggregated matrix $\mathbf {A}_{agg}$ and vector $\mathbf {y}_{agg}$ as follows:

    \begin{equation}
        \begin{aligned}
            \mathbf {A}_{agg} =\mathbf {A}_1||\mathbf {A}_2||\dots||\mathbf {A}_K\\
        \mathbf {y}_{agg}  =\mathbf {y}_1 ||\mathbf {y}_2|| \dots ||\mathbf {y}_K.  
        \end{aligned}
      \label{generate_Y}
    \end{equation}
    
    \item[3. ]The server computes \textbf{the hash watermark} $\mathbf{h}$ adopted to mark the trained FedDNN model:
     \begin{equation}
        \mathbf h=\mathtt H(\mathbf{A}_{agg}||\mathbf y_{agg}), 
        \label{generate_hash}
    \end{equation}

    in which $\mathtt H(\cdot)$ is a hash function $\mathtt{H}(\cdot):\{0,1\}^{*}\rightarrow\{0,1\}^{n}$.
    
    Finally, $\mathbf h, \mathbf{A}_{agg}, \mathbf{y}_{agg}$ are distributed to all clients as the watermark by the server.
    

    \item[4. ] The clients check $\mathbf h, \mathbf{A}_{agg}, \mathbf{y}_{agg}$ as follow. a) Each client checks whether $\mathbf{h}=\mathtt{H}(\mathbf{A}_{agg}||\mathbf{y}_{agg})$ to verify the correctness of the watermark. b) They check whether the $\mathbf{y}_{agg}$ and $\mathbf{A}_{agg}$ contain their public inputs to make sure the watermark $\mathbf{h}$ has their own ownership information. c) They check if there are $K$ public inputs in $\mathbf{y}_{agg}, \mathbf{A}_{agg}$ to ensure that there is no public input of the server in the watermark. If all the checks pass, then the server honestly performs watermark generation. The clients start training and watermark embedding. Else, the watermark generation will fail due to the dishonest behavior of the server.
\end{itemize}

\subsection{Watermark Embedding}
\label{watermark_embedding}
Following \cite{fan2021deepip,li2022fedipr}, each client embeds watermarks $\mathbf h$ into the Batch Normalization (BN) layer of the neural network. First, the scaling parameters of normalization weights $W_{\gamma}=(\gamma^{(1)},\cdots,\gamma^{(\omega)})$ (defined in Eq. \eqref{Bn_layer}) with $\omega$ channels is chosen according to the position parameter $\mathcal P$.
\begin{equation}
    \begin{split}
    \mathbf y^{(i)}=\mathbf \gamma ^{(i)} \ast \mathbf x^{(i)} +\mathbf\beta^{(i)},
    \end{split}
    \label{Bn_layer}
\end{equation}
where $\mathbf \gamma^{(i)}$ and $\mathbf\beta^{(i)}$ are the scaling and bias parameters in channel $i$ for BN layer respectively,  $\mathbf x^{(i)} $ is the input of the BN layer.

Then we add the following regularization term $\mathtt{L}_h$ (Eq. \eqref{hash_loss}) into the main task loss such that the sign of $\mathbf W_N \mathbf E$ is consistent with the watermarks $\mathbf h$:
\begin{equation}
    \begin{split}
    \mathtt{L}_{h}(\mathbf {W_\gamma},  \mathbf E, \mathbf h)&=\mathtt {HL}(\mathtt{sgn}(\mathbf W_\gamma  \mathbf E), \mathbf h)\\
    &=\mathtt {HL}(\mathbf h', \mathbf h)\\
    &=\sum\limits^n\limits_{i=1}max(\mu - b_i t_i, 0),
    \end{split}
    \label{hash_loss}
\end{equation}
where $\mathbf \mathbf h' = \mathtt{sgn}(\mathbf W  \mathbf E)=(b_1,\cdots,b_n)\in \{0,1\}^n$
is the extracted watermarks and $ \mathbf h=\mathtt H(\mathbf y)=(t_1,\cdots,t_n) \in \{0,1\}^n $ is the target watermark,  $\mathtt{HL ()}$ defines the Hinge loss \cite{rosasco2004loss} and $\mu$ is the parameter of hinge loss.

\subsection{Ownership Verification}
\label{ownership_verification}
In the ideal case, the xLPN public input is embedded directly into the watermark. The watermark is not damaged and the verifier can correctly obtain the original xLPN public input embedded into the model and perform a direct zero-knowledge ownership verification. In practice, it would introduce error $e = e_1 + e_2$ to the watermark, where $e_1$ is caused by normal model training \cite{uchida2017embedding} and $e_2$ is caused by watermark removal attack \cite{li2022fedipr}.

The presence of error makes it difficult to obtain the original xLPN public input from the watermark, which causes several security problems. a) The client may not be able to use the original xLPN public input to prove ownership. b) An attacker may change the original xLPN public input to his own constructed xLPN public input and pass the ownership verification. c) Error on the xLPN public input can cause difficulties in proving security.

To solve these issues, we split ownership verification into two steps: validity check and zero-knowledge ownership verification. The validity check is used to address security issues caused by errors of embedding watermarks. In the zero-knowledge verification, the client uses the xLPN private input i.e., the credential to prove that he embeds a watermark on the federated model and thus declares ownership of the model $\mathbb{M}[\mathbf W,\mathbf {h'}]$. Noted that the credentials are not leaked to the verifier for our zero-knowledge ownership verification scheme, which doesn’t need a trusted verifier to guarantee the confidentiality of credentials.

In the following, we call the entity attempting to prove ownership a prover, which may be either a client or an attacker.

\subsubsection{Validity Check}
\label{Validity Check}\
First, the verifier gets the watermark $\mathbf{h}'$ from the model $\mathbb{M}[\mathbf{W},\mathbf{h}']$. Then the prover sends the aggregated public xLPN input $\mathbf{A}_{agg}, \mathbf{y}_{agg}$ to the verifier. If the Hamming distance between $\mathbf h'$ and $\mathtt{H}(\mathbf{A}_{agg}||\mathbf{y}_{agg})$ is less than $err_n$, the verifier view $\mathbf{A}_{agg}$ and $\mathbf{y}_{agg})$ as valid aggregated xLPN public input. Finally, the prover selects $\mathbf{A}$ from $\mathbf{A}_{agg}$ and $\mathbf{y}$ from $\mathbf{y}_{agg}$, and uses $\mathbf{A}$ and $\mathbf{y}$ to perform zero-knowledge ownership verification.

For issue a) in Sect. \ref{ownership_verification}, provers who have the right xLPN public input will pass the validity check (in Sect. \ref{Completeness}). Sect; For attackers who don't own the origin aggregated xLPN public input, the probability of an attacker finding the valid one is negligible (in Sect. \ref{Scheme Security}). Moreover, compared to analyzing the error distribution, we can make a simpler security proof from the number of bits that the watermark is damaged.

\subsubsection{Zero-knowledge Ownership Verification}
\label{Sect:OwnershipVerification}
In zero-knowledge ownership verification, the credential of the prover is not leaked to the verifier. So FedZKP doesn't need a trusted verifier to guarantee the confidentiality of credentials, which can't be achieved by existing FMOV schemes.

The public input for the prover and verifier is a matrix $\mathbf{A}$ and a vector $\mathbf{y}$, where $\mathbf y = \mathbf{A}\cdot\mathbf{s}\oplus \mathbf{e}$. The private input of the prover is two vectors $\mathbf{e},\mathbf{s}$. The commitment scheme can be anyone which is perfect binding. Here we use the commitment scheme \cite{jain2012commitments} based on the xLPN problem. To achieve a tiny-enough knowledge error, the zero-knowledge proof can be repeated $d$ rounds. Each round of the protocol can be executed as follows:
\begin{itemize}
    \item The prover samples a random permutation $\pi$. Then, he selects two vectors $\mathbf{v}\stackrel{R}{\leftarrow}\{0,1\}^{l}, \mathbf{f}\stackrel{R}{\leftarrow}\{0,1\}^{m}$, and then sends the commitments $C_0, C_1, C_2$ to the verifier, where
    \begin{equation}\nonumber
        \begin{aligned}
            &C_0=\mathtt{Com}(\pi, \mathbf{t}_0=\mathbf{A}\cdot\mathbf{v}\oplus\mathbf{f})\\
            &C_1=\mathtt{Com}(\mathbf{t}_1=\pi(\mathbf{f}))\\
            &C_2=\mathtt{Com}(\mathbf{t}_2=\pi(\mathbf{f}\oplus\mathbf{e})).
        \end{aligned}
    \end{equation}
    \item The verifier selects a random number $c\stackrel{R}{\leftarrow}\{0,1,2\}$ and sends it to the prover.
    \item Depending on the value of random number $c$, the prover opens the following commitments:
    \begin{itemize}
        \item[$c=$0:] The prover opens the commitments $C_0,C_1$ by sending the random permutation $\pi$ and the vectors $\mathbf{t}_0,\mathbf{t}_1$ and the associated random openings.
        \item[$c=$1:] The prover opens the commitments $C_0,C_2$ by sending the random permutation $\pi$ and the vectors $\mathbf{t}_0,\mathbf{t}_2$ and the associated random openings.
        \item[$c=$2:] The prover opens the commitments $C_1,C_2$ by sending the vectors $\mathbf{t}_1,\mathbf{t}_2$ and the associated random openings.
    \end{itemize}
    \item The verifier verifies the correctness of the openings received from the prover, and additionally performs the following checks depending on the challenge $c$:
    \begin{itemize}
        \item[$c=$0:] The verifier checks whether $\mathbf{t}_0\oplus\pi^{-1}(\mathbf{t}_1)\in\text{img}\ \mathbf{A}$.
        \item[$c=$1:] The verifier checks whether $\mathbf{t}_0\oplus\pi^{-1}(\mathbf{t}_2)\oplus\mathbf{y}\in\text{img}\ \mathbf{A}$.
        \item[$c=$2:] The verifier checks whether the Hamming weight of $\mathbf{t}_1\oplus \mathbf{t}_2$ is $w$.
    \end{itemize}
\end{itemize}

The vector $\mathbf{a}\in\text{img}\ \mathbf{A}$ means that there exist a vetor $\mathbf{b}$ so that $\mathbf{A}\cdot\mathbf{a}=\mathbf{b}$. If any of the $d$ rounds fails, the prover fails to pass ownership verification. If all $d$ rounds pass, the prover passes the ownership verification. The knowledge error in single round xLPN ZKP is $\frac 2 3$ \cite{jain2012commitments}, with $d$ increases, the knowledge error decreases exponentially to $(\frac{2}{3})^{d}$. The security of the scheme is discussed in detail in Sect. \ref{Security Analysis}.
\section{Security Analysis}
\label{Security Analysis}
In this section, first, we introduced the definition of \emph{security boundary}. Second, we prove the \emph{completeness} and \emph{honest-verifier zero-knowledge} of FedZKP. After that, we give the \emph{security model} of FedZKP. Finally, we give the \emph{security proof} that it is difficult for an attacker to successfully pass the ownership verification of a model that does not belong to him. We denote ${a \choose b}$ as the combination number.

\subsection{Security Boundary}
\label{Sect:Security Boundary}
We give the definition of the security boundary. However, to facilitate understanding, we need to first give the probability of finding a hash near collision.

Given a binary string $\mathbf{h}\in\{0,1\}^{n}$ output from hash function, the size of the $n'$-near-collision subset for $\mathbf{h}$ is $\sum^{n'}_{i=0}{i \choose n}$, the probability $p$ of the adversary that finding a $n'$-near collision to $\mathbf{h}$ from a single computation of the hash is 
\begin{equation}
\label{near-collision probability}
    p=2^{-n}\cdot\sum^{n'}_{i=0}{i \choose n}.
\end{equation}

\begin{myDef}[\textbf{Watermark Detection Errors $r$}] \label{def:Watermark Detection Errors}
For a $n$ bits targeted watermark $\mathbf{h}$ and a given model $\mathbb {M}[\mathbf{W},\mathbf{h'}]$, the watermark detection rate is calculated as:
    \begin{equation}
        r =\mathtt{Dif} (\mathbf{h},\mathbf{h'})/n
    \end{equation}
    in which $\mathtt{Dif}(\mathbf{h},\mathbf{h'})$ is the Hamming distance between $\mathbf{h}$ and $\mathbf{h'}$.     
\end{myDef}


\begin{myDef}[\textbf{Security Boundary of Watermark Detection Errors $err_n$}]
\label{Def:Security Boundary}
Given the watermark length $n$ and the required probability $p_r$ of finding near collision, the maximum value of the number of bits $err_n$ that an attacker can corrupt to pass the validity check is the security boundary. The relationship between the security boundary $err_n$, the probability $p_r$ and the watermark length $n$ satisfies the following equation:
\begin{equation}
    \begin{aligned}
   \frac{1}{2^n} \sum \limits_{i=0}\limits^{ 2err_n-1 }{i \choose n}< p_r&\leq\frac{1}{2^n} \sum \limits_{i=0}\limits^{ 2err_n }{i \choose n}.
    \end{aligned}
\end{equation}
Moreover, given the watermark length $n$, we can derive the watermark detection rate $r_n$ required for an attacker to find a near-collision probability less than $p_r$:
\begin{equation} \label{eq:lower-bound-r}
    r_n = 1-err_n /n.
\end{equation}
\end{myDef}

We assume that the number of bits that an attacker corrupts the watermark is less than the security boundary $err_n$ in the security analysis. We set $p_r=1/2^{128}$ in the experiment in Sect. \ref{Roboustness}. The experimental results show that without unduly degrading the accuracy of the model, the number of corrupted bits of watermark caused by the existing removal attack is less than the security boundary $err_n$. So the assumption is reasonable.

\subsection{Completeness}
\label{Completeness}
Completeness means that the client can always pass ownership verification. We assume that the attacker corrupts watermark bits that are always less than the security boundary $err_n$. The client can always give the aggregated xLPN public input $\mathbf A_{agg}$ and $\mathbf y_{agg}$ so that the Hamming distance between its hash value and the extracted watermark is less than $err_n$, and pass the validity check. Due to the completeness of xLPN ZKP \cite{jain2012commitments}, the provers with private input always pass the zero-knowledge proof. So the client with private input i.e., credential can always pass the ownership verification. In summary, FedZKP has completeness.

\subsection{Honest-Verifier Zero-Knowledge}
\label{Honest-Verifier Zero-Knowledge}
Honest-verifier zero-knowledge means that the verifier cannot get any information about the private input during the proof interaction. To prove this property, we discourse that there exists a simulator $\mathcal{S}$ without xLPN private input, and that the verifier cannot distinguish between the simulator $\mathcal{S}$ and the client with xLPN private input. This illustrates that the verifier does not get any information about the xLPN private input used as credentials from the proof interaction. The output of the simulator $\mathcal{S}$ varies with the random number $c\in\{0,1,2\}$ sent by the verifier. The simulator is constructed as follows:
\begin{itemize}
    \item[$c=$0:] Simulator $\mathcal{S}$ just computes the commitments $C_0, C_1$, which is similar to the output of the client. The commitment $C_2$ is the commitment of number 0. The distribution of $C_0, C_1,\pi',\mathbf{t}_0,\mathbf{t}_1$ is identical to that in the real protocol. Because the commitment function $\mathtt{Com}(\cdot)$ is computationally hiding, the distribution of $C_2$ is computationally indistinguishable from the real protocol with an honest client. So $\mathcal{S}$ and client are indistinguishable when the random number $c$ is 0.
    \item[$c=$1:] Simulator $\mathcal{S}$ generates random permutation $\pi$, and selects vectors $ \mathbf{a}\stackrel{R}{\leftarrow}\{0,1\}^{m}$ and $\mathbf{b}\stackrel{R}{\leftarrow}\{0,1\}^{l}$. It computes commitments $C_0=\mathtt{Com}(\pi,\mathbf{A\cdot b\oplus y\oplus a}), C_2=\mathtt{Com}(\pi(\mathbf{a}))$. Then $C_1$ is computed as the commitments of number 0. In the simulation, vectors $\mathbf t_0, \mathbf t_2$ is computed as $\mathbf{t}_0=\mathbf{A}\cdot \mathbf{b}\oplus \mathbf{y}\oplus \mathbf{a}, \mathbf{t}_2=\pi(\mathbf{a})$. The distribution of $\mathbf{t}_2$ in the real protocol is uniform. When the verifier opens $C_0$, vector $\mathbf{t}_0$ equals $\mathbf{A\cdot v\oplus f}$ in the real protocol; the vector $\mathbf{t}_0$ equals $\mathbf{A}\cdot(\mathbf{b\oplus s})\oplus(\mathbf{a}\oplus\mathbf{e})$ in the simulation. Obviously, vectors $\mathbf{v}$ and $\mathbf{b}\oplus\mathbf{s}$ are both uniformly random, then the distribution of vectors $\mathbf{f}$ and $\mathbf{a\oplus e}$ is equal. And the distribution of $C_1$ is computationally indistinguishable because the commitment function $\mathtt{Com}$ is computationally hiding. Since the distribution of the parameters sent by the simulator $\mathcal S$ and the parameters sent by the client is the same when $c=1$, simulator $\mathcal{S}$ and client are indistinguishable when the random number $c$ is 1.
    \item[$c=$2:] $\mathcal{S}$ selects $\mathbf{a}\stackrel{R}{\leftarrow}{\{0,1\}^{m}}$ and $\mathbf{b}\stackrel{R}{\leftarrow}{\{0,1\}^{m}_w}$. Then it computes $C_0=\mathtt{Com}(0), C_1=\mathtt{Com}(\mathbf{a}), C_2=\mathtt{Com}(\mathbf{a}\oplus\mathbf{b})$. The $\mathbf{a}$ and $\mathbf{a\oplus b}$ are the random number unique to the output of random permutation $\pi$. And $C_0$ is computationally indistinguishable by the hiding property of the commitment function $\mathtt{Com}(\cdot)$. So $\mathcal{S}$ and client are indistinguishable when the random number $c$ is 2.
\end{itemize}

In summary, the honest-verifier zero-knowledge of xLPN ZKP is proven. The verifier does not have access to any information about the credential during the verification process, which effectively avoids the problem of credential leakage.

\subsection{Security Model}
\label{Sect:SecurityModel}
Here we design the security model for FedZKP. We define the information available to an attacker in a real-world scenario and the conditions for a successful attack and construct the security model of FedZKP. For convenience, we assume that one watermark corresponds to one xLPN instance, rather than one watermark corresponding to a polynomial number of xLPN instances in the real scheme. Since the number of watermarks and instances obtained by the attacker in both cases mentioned above are at a polynomial level, this assumption doesn't impact the security proof. The detailed capabilities of the attacker are shown as follows:
\begin{enumerate}
     \item [Cap. 1] Obtaining the watermarks, public inputs, commitments, random numbers, and the corresponding openings for the commitments sent by the clients xLPN ZKP for ownership verification.
    \item [Cap. 2] Stealing some models and destroying the watermark with removal attack with a removal attack on the premise that the accuracy of the model is reduced acceptably. Here the attacker can also get the watermarks of the stolen models.
    \item [Cap. 3] Generating own xLPN instances and trying to find one that the Hamming distance between hash values and the watermarks obtained in the above two points is less than $err_n$.
\end{enumerate}


We enhance the attacker's capabilities to combine Cap. 1 and Cap. 2 to make the security model more concise. The attacker gets $q$ hash watermarks of the model and the corresponding parameters including public inputs, commitments, random numbers, and openings in ZKP. And he can destroy up to $err_n$ bits of the watermark.

The attacker using his ability can pass the verification in the following way:
\begin{enumerate}
    \item [1. ] The attacker replaces the xLPN public input in the watermark. the attacker first finds a public where the Hamming distance between its hash value and the origin watermark is less than $2err_n$. And then damages the $err_n$-bit of the watermark according to the hash value of public input. From the verifier's point of view, the Hamming distance between the hash of the public input and the corrupted watermark is less than $err_n$, which is considered valid public input. Finally, the attacker proves to the verifier that he has the private input for the public input to pass ownership verification.
    \item [2.] The attacker launches the brute force attack when the attacker does not find a $2err_n$-near collision. It's still available for the attacker to perform ZKP for above $q$ undamaged watermarks by using the xLPN public inputs he gets. If he passes any of the ZKP of the $q$ public inputs, he still managed to achieve his target.
\end{enumerate}

Based on the above analysis, we formalize the attacker's capabilities and goals as the security model. The security model is a formal game of interaction between two entities, the challenger $\mathcal{C}$ and the adversary $\mathcal{A}$. The challenger has xLPN private input and the adversary can be viewed as a collection of attacks in reality.

\begin{enumerate}
    \item [1. ] $Setup:$ The Challenger $\mathcal{C}$ generates the public key $pk$ which contains
    \begin{equation}\nonumber
        \begin{aligned}
            \bigg(0\leq\tau\leq\frac{1}{2}, w, \mathtt{Com}(\cdot),\mathtt{H}(\cdot):\{0,1\}^{*}\rightarrow\{0,1\}^{n}, err_n\bigg).
        \end{aligned}
    \end{equation}
    \item [2. ] $Challenge\ Instances\ Gen:$ The Challenger $\mathcal{C}$ generates $q$ xLPN instances $(\mathbf{A}_1,\mathbf{s}_1,\mathbf{e}_1,\mathbf{y}_1=\mathbf{A}_1\cdot\mathbf{s}_1\oplus\mathbf{e}_1),...,(\mathbf{A}_q,\mathbf{s}_q,\mathbf{e}_q,\mathbf{y}_k=\mathbf{A}_q\cdot\mathbf{s}_q\oplus\mathbf{e}_q)$, and computes the challenge hash watermarks $\mathbf{h}_1=\mathtt{H}(\mathbf{A}_1||\mathbf{y}_1),...,\mathbf{h}_q=\mathtt{H}(\mathbf{A}_q||\mathbf{y}_q)$. Then the challenger $\mathcal{C}$ sends challenge public inputs $(\mathbf{A}_1,\mathbf{y}_1)$ ,..., $(\mathbf{A}_q,\mathbf{y}_q)$ and challenge hash watermarks $\mathbf{h}_1,...,\mathbf{h}_q$ to the adversary $\mathcal{A}$.
    \item [3. ] $Phase:$ The Adversary $\mathcal{A}$ chooses a public input $(\mathbf{A}_i,\mathbf{y}_i)$ and hash watermark $\mathbf{h}_i$, $i\in\{1,...,q\}$ and asks the challenger $\mathcal{C}$ to act as the prover in ZKP to interact with the adversary $\mathcal{A}$. The adversary $\mathcal{A}$ records the parameters in the interaction. The $Phase$ can be repeated polynomial times.
    \item [4. ] $Instances\ Gen:$ The adversary $\mathcal{A}$ generates $k$ xLPN$_\tau$ instances $(\mathbf{A}'_1,\mathbf{s}'_1,\mathbf{e}'_1,\mathbf{y}'_1=\mathbf{A}'_1\cdot\mathbf{s}'_1\oplus\mathbf{e}'_1),...,(\mathbf{A}'_k,\mathbf{s}'_k,\mathbf{e}'_k,\mathbf{y}'_k=\mathbf{A}'_k\cdot\mathbf{s}'_k\oplus\mathbf{e}'_k)$. 
    \item [5. ] $Challenge\ 1:$ The adversary $\mathcal{A}$ sends an xLPN instance $(\mathbf{A}'_j,\mathbf{s}'_j,\mathbf{e}'_j,\mathbf{y}'_j),j\in\{1,...,k\}$ to the challenger $\mathcal{C}$. If $(\mathbf{y}'_j=\mathbf{A}'_j\cdot\mathbf{s}'_j\oplus\mathbf{e}'_j)$, $||\mathbf{e}'_j||_1=\lfloor k\tau \rceil$ and exists $i\in\{1,...,q\}$ that the Hamming distance between $\mathtt{H}(\mathbf{A}'_j||\mathbf{y}'_j)$ and $\mathbf{h}_i$ is less than the security boundary $err_n$, the adversary $\mathcal{A}$ wins the game.
    \item [6. ] $Challenge\ 2:$ If the adversary $\mathcal{A}$ doesn't win in $Challenge1$, then he performs as the prover on every challenge public input with the challenger $\mathcal{C}$. If the adversary $\mathcal{A}$ passes the ZKP of any of the public inputs, he wins the game.
\end{enumerate}

The adversary $\mathcal{A}$ winning in a secure model is equivalent to an attacker successfully declaring ownership of a model that does not belong to him in the real world. $E_1$ is the event that the adversary $\mathcal{A}$ wins in $Challenge1$. $\neg E_1$ is the event that the adversary $\mathcal{A}$ doesn't win in $Challenge1$. $E_2$ is the event that $\mathcal{A}$ wins in $Challenge2$. The advantage of $\mathcal{A}$ is the probability that $\mathcal{A}$ wins the game:
\begin{equation}\nonumber
    \begin{aligned}
        Adv(\mathcal{A})=Pr[E_1]+Pr[E_2|\neg E_1]
    \end{aligned}
\end{equation}

\begin{myDef}
    \label{Def:Attacker}
     If the advantages of winning the above security model are negligible for all PPT adversaries. It is very difficult for an attacker in FedZKP to successfully pass ownership verification of a model that does not belong to the attacker.
\end{myDef}

\subsection{Scheme Security}
\label{Scheme Security}
We assume that the number of watermarked bits corrupted by the attacker is less than the security bound defined in Def. \ref{Def:Security Boundary}. We prove that the adversary's advantage in winning the security model is negligible based on this assumption. The random oracle queries in Theo. \ref{Advantage Theorem} represents the attacker computes the hash function in reality.

\begin{myTheo}
\label{Advantage Theorem}
Suppose the hash function $\mathtt{H}(\cdot)$ is a random oracle \cite{bellare1993random}. $n$ is the output length of the random oracle. $q$ is the number of xLPN challenge public inputs that the adversary obtains. $k$ is the max time for the adversary to query the random oracle. and the ZKP of each xLPN public input requires $d$-rounds repetition. The number of bits of the attacker's corrupted watermark is less than $err_n$. The advantage that for any PPT adversary $\mathcal{A}$ holds that:
\end{myTheo}

\begin{equation}
    \begin{aligned}
        Adv(\mathcal{A})\leq \frac{kq}{2^{n}}\cdot\sum^{2err_n}_{i=0}{i \choose n}+q(\frac{2}{3})^{d}
    \end{aligned}
\end{equation}

\noindent \textbf{Proof}.

After the adversary $\mathcal{A}$ gets $q$ challenge hash watermarks $\mathbf{h}_1,...,\mathbf{h}_q\in\{0,1\}^{n}$. $\mathcal{A}$ sends a single query to the random oracle, which returns a hash value $\mathbf{h}\in\{0,1\}^{n}$ and the probability $p_1$ that no $2err_n$-near collisions happen for hash value $\mathbf{h}$ to all challenge hash watermarks satisfies
\begin{equation}
\label{SingleNearCollision}
    \begin{aligned}
        p_1&\geq1-q\cdot2^{-n}\cdot\sum^{2err_n}_{i=0}{i \choose n}
    \end{aligned}
\end{equation}

In Eq. (\eqref{SingleNearCollision}), if there is no intersection between the $2err_n$-near collision subsets of all challenge hash watermarks $\mathbf{h}_1,...,\mathbf{h}_q\in\{0,1\}^{n}$, $p_1=1-q\cdot2^{-n}\cdot\sum^{2err_n}_{i=0}{i \choose n}$. Else the concentration size of the $2err_n$-near subsets for all challenge hash watermarks will become smaller and the probability $p_1$ is larger than $1-q\cdot2^{-n}\cdot\sum^{2err_n}_{i=0}{i \choose n}$. 

\begin{figure}
    \centering
    \includegraphics[scale=0.13]{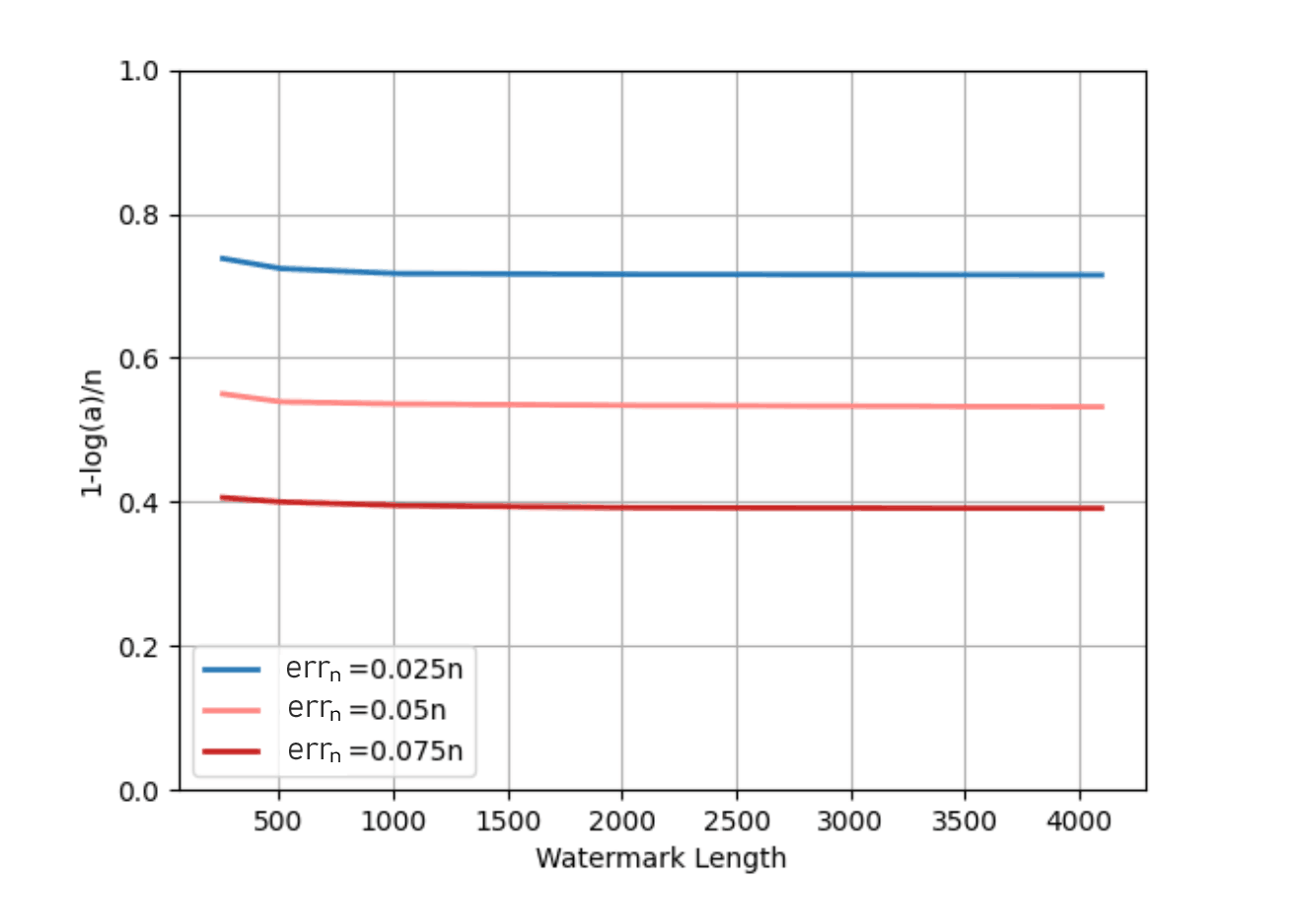}
    \caption{This figure illustrates that the value of $1-log(a)/n$ converges as watermark length $n$ increases.}
        \label{Fig:log}
\end{figure}

When the adversary $\mathcal{A}$ asks the oracle $k$ times, the probability $p_2$ that the oracle still does not output a $2err_n$-near-collision for all challenge hash watermarks satisfies
\begin{equation}
\label{KTimesNearCollision}
    \begin{aligned}
        p_2&\geq(1-q\cdot2^{-n}\cdot\sum^{2err_n}_{i=0}{i \choose n})^k\\
        &\geq 1-kq\cdot2^{-n}\cdot\sum^{2err_n}_{i=0}{i \choose n}
    \end{aligned}
\end{equation}

The probability $p_3$ is that the adversary $\mathcal{A}$ gets any $2err_n$-near-collision to xLPN challenge public inputs (i.e., the event $E_1$ happens) for $q$ challenge hash watermarks. is less than $kq\cdot2^{-n}\cdot\sum^{2err_n}_{i=0}{i \choose n}$.

The knowledge error of the xLPN ZKP is $\frac{2}{3}$ \cite{jain2012commitments}. The probability that $\mathcal{A}$ repeats a $d$-round ZKP on a single xLPN challenge public input and passes is $(\frac{2}{3})^d$. The probability $p_4$ that $\mathcal{A}$ passes any verification of the $q$ xLPN challenge public inputs (i.e., event $E_2$ happens) is
\begin{equation}
    \begin{aligned}
        p_4&=1-\big(1-(\frac{2}{3})^d\big)^{q}\\
        &\leq q\cdot(\frac{2}{3})^d
    \end{aligned}
\end{equation}

The advantage of $\mathcal{A}$ is
\begin{equation}
    \begin{aligned}
        Adv(\mathcal{A})&=Pr[E_1]+Pr[E_2|\neg E_1]\\
        &\leq Pr[E_1]+Pr[E_2]\\
        &\leq kq\cdot2^{-n}\cdot\sum^{2err_n}_{i=0}{i \choose n}+q(\frac{2}{3})^d
    \end{aligned}
\end{equation}

Let $a=\sum^{2err_n}_{i=0}{i \choose n}$. Fig. \ref{Fig:log} illustrates that the value of $1-log(a)/n$ tends to converge as the watermark length $n$ grows for the same $err_n$. When $err_n=\lfloor0.025n\rfloor$, $1-log(a)/n$ converges to 0.713. So $a/2^n=1/2^{0.713n}$, the advantage of the adversary is $kq\cdot\frac{1}{2^{0.713n}}+q(\frac{2}{3})^d$. When $err_n=\lfloor0.075n\rfloor$, $1-log(a)/n$ converges to 0.39, and $a/2^n=1/2^{0.39n}$. The advantage of the adversary is $kq\cdot\frac{1}{2^{0.39n}}+q(\frac{2}{3})^d$.

Since $k,q$ are polynomial sizes which are negligible compared to $2^{0.713n}$, $2^{0.39n}$, and $(\frac{2}{3})^{d}$, the advantage that the adversary $\mathcal{A}$ wins in the security model is negligible. According to the above conclusion and Def. \ref{Def:Attacker}, it is very difficult for the attacker to successfully pass ownership verification of a model that does not belong to him. The security of FedZKP is guaranteed for any attack that can be captured by the security model, including potential attacks.
\section{Experimental Analysis} 
\label{Sect:Experimental Analysis}
This section illustrates the empirical study of the proposed FedZKP with a focus on the \textit{fidelity, robustness in federated training, robustness to removal attacks} of watermark embedding, and \textit{time cost} of zero-knowledge ownership verification. 

\subsection{Experimental Setup}

\textit{Models}: We choose the well-known deep neural networks AlexNet \cite{krizhevsky2017imagenet} and Resnet18 \cite{he2016deep} as experimental models. We embed the watermark into multiple normalization scale weights in the network.

\noindent \textit{Datasets}: We choose the classic public image sets CIFAR10 and CIFAR100 \cite{krizhevsky2009learning} as the datasets for model classification tasks because these public datasets are conducive to comparing our scheme with the previous work.

\noindent \textit{Federated Learning Settings}: In each training round, each client uses its own data to embed the watermark and update the model locally, and then uploads parameters to the server. The server uses the mainstream FedAvg algorithm \cite{mcmahan2017communication} to aggregate the received parameters and returns aggregated results to each client. If no additional instructions are given, the number of FL clients is 10, and the original datasets are evenly distributed to each client. The hash function is implemented by SHAKE algorithm \cite{dworkin2015sha} of SHA3. The embedding matrix is generated from the standard normal distribution, i.e., $\mathbf E^{\omega \times n} \stackrel{R}{\leftarrow}\mathcal{N}(0,1)$.
Other detailed FL settings can be seen in Appendix \ref{Sect:AppendixExp}

\noindent \textit{ZKP Settings}: In the experiment of zero-knowledge ownership verification, we randomly select the xLPN error vector from the Bernoulli distribution with parameter $\tau=1/4$. We evaluate the time, memory, and communication cost for xLPN zero-knowledge proofs and commitments for different numbers of columns $l$. The selection of parameters can be referred to in some literature \cite{levieil2006improved,pietrzak2012cryptography}. If xLPN wants to achieve 80 bits security, it's recommended \cite{levieil2006improved} using $l = 512, \tau = 1/4$.

\subsection{Evaluation Metrics} \label{sec:eva}

For model training and watermark embedding, we evaluate FedZKP from three aspects: fidelity, robustness, and embedded effect.  
For zero-knowledge ownership verification, its security is mainly proved by security analysis in Sect. \ref{Security Analysis}, and here we mainly verify that the watermark embedding effect of FedZKP meets the security requirements of zero-knowledge verification.

\noindent\textbf{Fidelity:} A good watermarking scheme should preserve the model performance, so we use the \textit{main task accuracy} of the model as the fidelity evaluation index.


\noindent\textbf {Robustness} A good watermark should be preserved as much as possible in the FL training or against various attacks, so we choose the \textit{watermark detection rate} $r$ as the robustness evaluation index. A higher watermark detection rate indicates the ownership scheme is more robust. 
\begin{myDef}[\textbf{{Watermark Detection Rate $r$}}] \label{def:watermark-detect}
For a $n$ bits targeted hash watermark $\mathbf{h}$ and a given model $\mathbb {N}[\mathbf{W},\mathbf{h'}]$, the watermark detection rate is calculated as:

    \begin{equation}
        \label{watermark_det_rate}
        r=1-\frac{1}{n}err
    \end{equation}
    in which $err$ is the watermark detection errors in Def. \ref{def:Watermark Detection Errors}. 
\end{myDef}
We provide a theoretical lower bound of the watermark detection rate $r_n$ in Eq. \eqref{eq:lower-bound-r}. The security of FedZKP is guaranteed if $r\geq r_n$.

\subsection{Fidelity}

To evaluate the utility of FedZKP watermarking, we compare the main task accuracy of  FedZKP with the baseline FedAvg \cite{mcmahan2017communication}. 

\begin{figure}[htbp]

    \centering
    \subfigure[AlexNet with CIFAR10]{
        \begin{minipage}[t]{0.5\linewidth}
        \centering
        \includegraphics[width=1.7in]{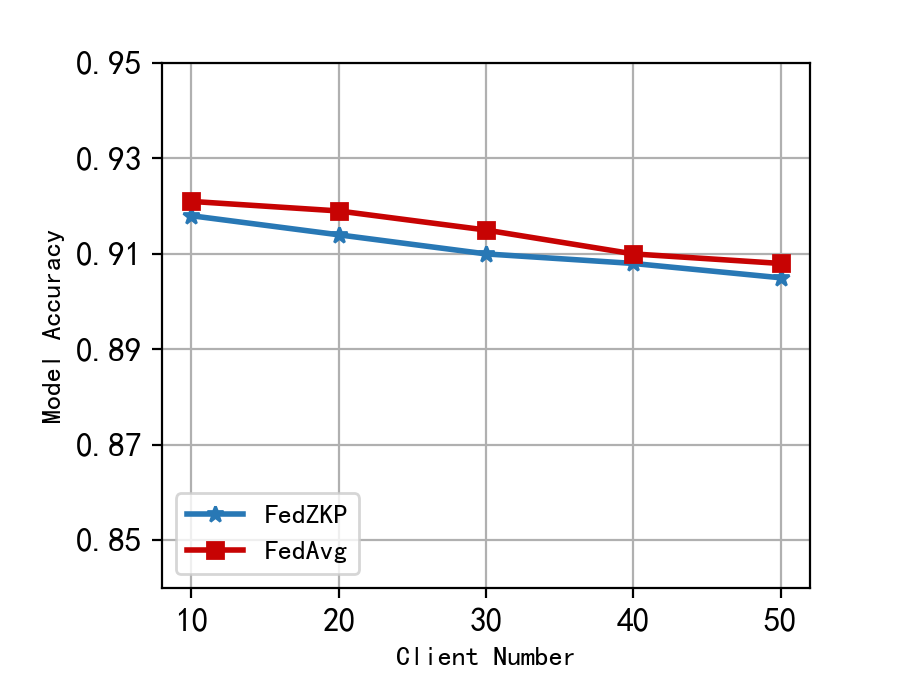}
        \end{minipage}%
    }%
    \subfigure[ResNet18 with CIFAR100]{
        \begin{minipage}[t]{0.50\linewidth}
        \centering
        \includegraphics[width=1.7in]{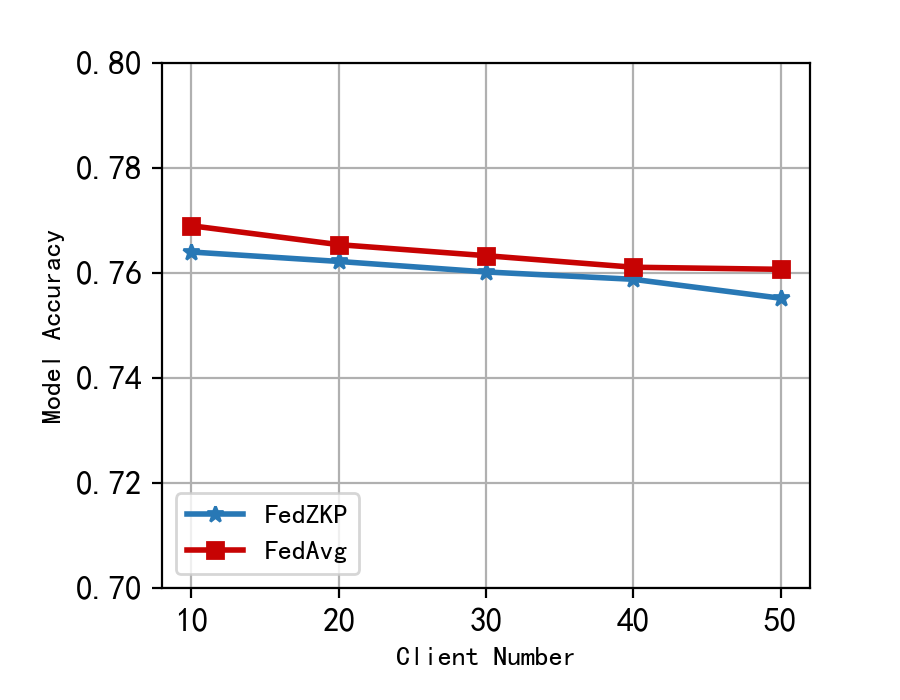}
        \end{minipage}%
    }%
    \centering
    \caption{Figure (a)-(b) illustrate the main task accuracy of the model in datasets with a varying number of clients (from 10 to 50).}
    \label{Fidelity}

\end{figure}
Fig. \ref{Fidelity} shows that the proposed FedZKP only causes an acceptable drop of main task accuracy within 2\%, which is the same as the FedIPR \cite{li2022fedipr}. This little drop is mainly due to the fact that taking watermark embedding as a secondary task during training limits the value space of model parameters.

\subsection{Robustness}
\label{Roboustness}
In this subsection, we demonstrate the robustness of FedZKP watermarking in two aspects: 1) whether FedZKP is robust against watermark removal attacks such as Fine-tuning, pruning, and Targeted Destruction Attack; 2) whether FedZKP is robust in the federated training process. Moreover, experimental results on the Robustness of FedZKP verify that the security boundary provided in Sect. \ref{Sect:Security Boundary} is reasonable. 
\begin{figure}[htbp]
    \centering
    \subfigure[Fine-tuning Attack]{
        \label{FT-attack}
        \begin{minipage}[t]{0.5\linewidth}
        \centering
        \includegraphics[width=1.7in]{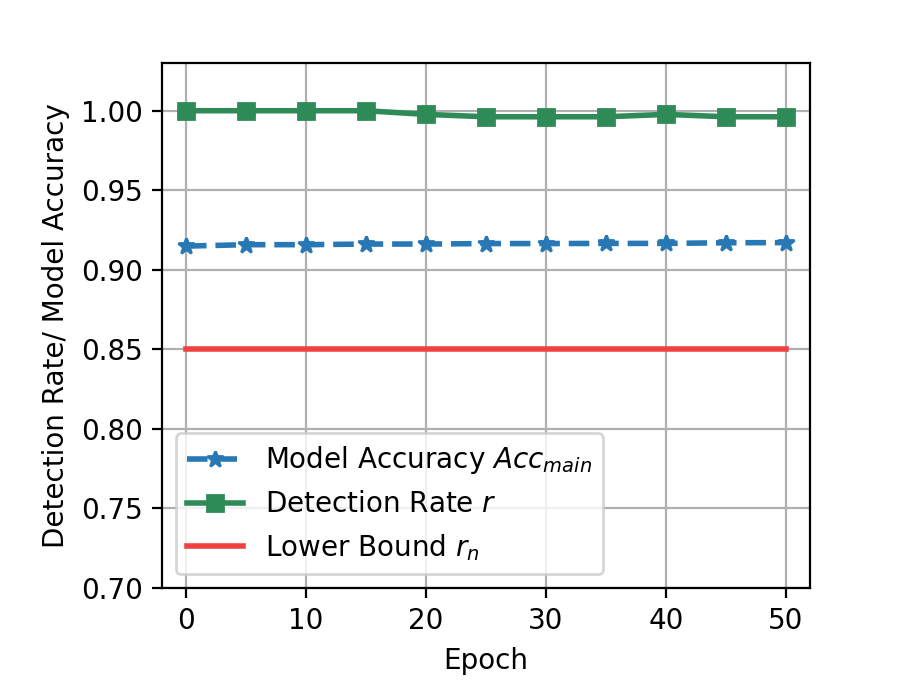}
        \end{minipage}%
    }%
    \subfigure[Pruning Attack]{
        \label{Purning-atttack}
        \begin{minipage}[t]{0.50\linewidth}
        \centering
        \includegraphics[width=1.7in]{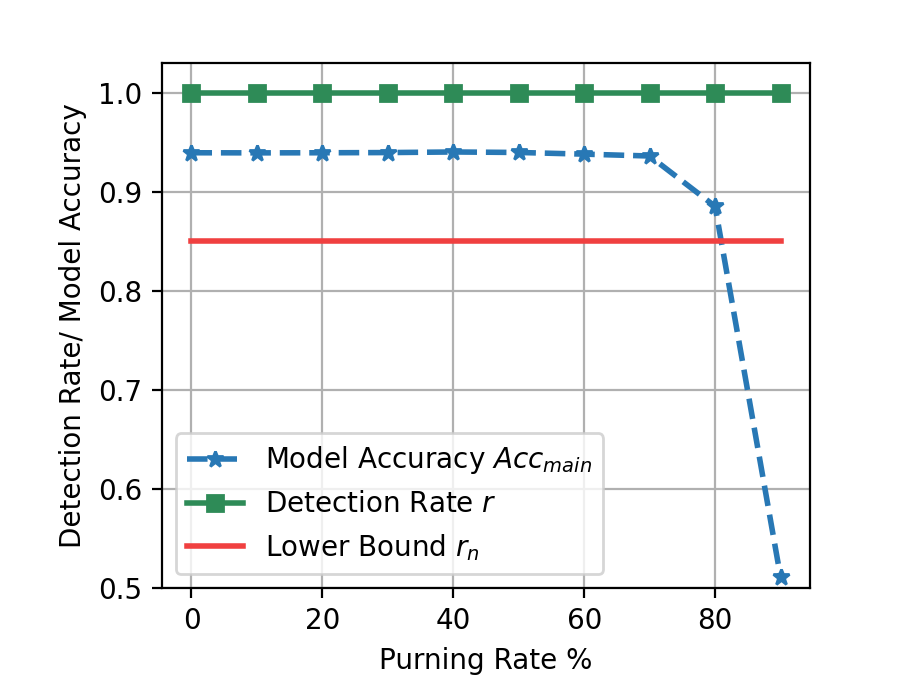}
        \end{minipage}%
    }%
    \centering
    \caption{Figure describes the robustness of our FedZKP under removal attacks. Figure (a) shows the robustness to fine-tuning of AlexNet with CIFAR10 and figure (b) shows the robustness to pruning of Resnet18 with CIFAR10.}
\end{figure}

\begin{figure*}
    \centering
    \subfigure[ResNet with 1024 bits Watermark]{
        \label{Target_Res_1024}
        \includegraphics[scale=0.39]{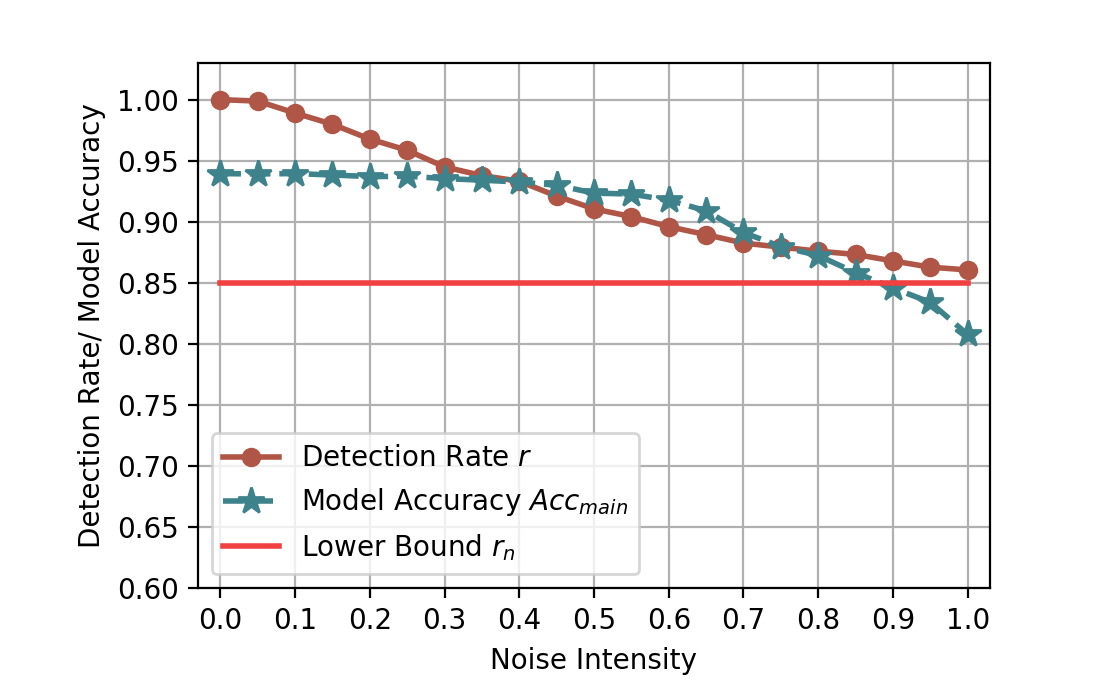} 
    }
    \subfigure[ResNet with 2048 bits Watermark]{
        \label{Target_Res_2048}
        \includegraphics[scale=0.39]{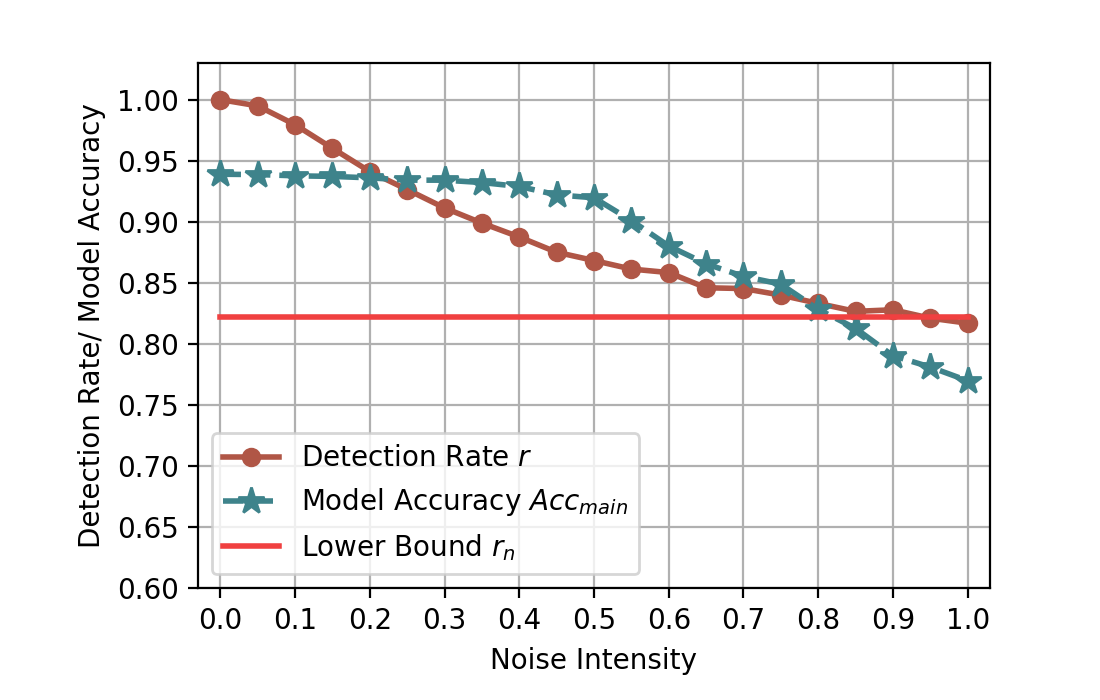} 
    }
    \subfigure[ResNet with 3076 bits Watermark]{
        \label{Target_Alex_1024}
        \includegraphics[scale=0.39]{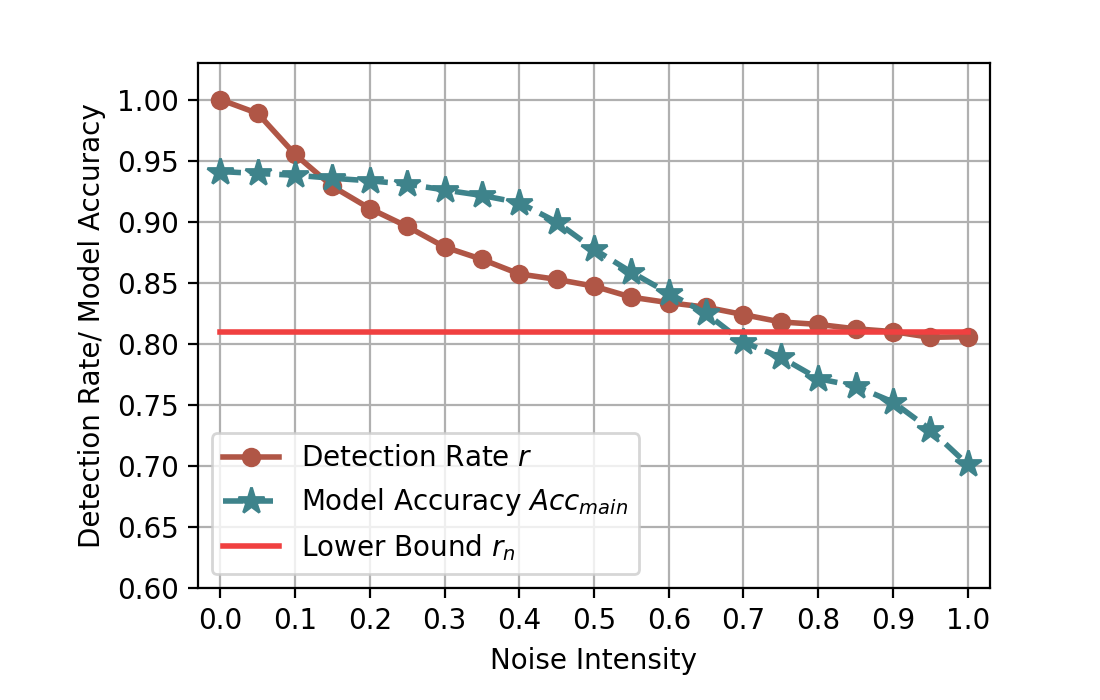}
    }
    \DeclareGraphicsExtensions.
    \caption{Figure (a)-(c) demonstrates the robustness of the FedZKP scheme against targeted destruction attacks on ResNet18 with 1024, 2048, and 3076 bits watermark, which illustrates that as the noise intensity increases, although there is some drop in watermark detection rate, the model performance suffers severely.}
    \label{targeted_attack}
\end{figure*}

\subsubsection{Robustness against Removal Attack}

Fine-tuning \cite{simonyan2014very} and pruning \cite{see2016compression} are two watermark removal techniques that do not require additional knowledge of the model and data. Their purpose is to remove or destroy the watermark contained in the model parameters as much as possible so that the watermark cannot be detected completely enough for ownership certification. 

\noindent\textbf{Robustness against Fine-tuning Attack:}
Fine-tuning attack on watermarks is conducted to train the network without the presence of the regularization term $\mathtt {L}_{\mathbf{h}}$ in Eq. \eqref{combined_loss}.
Fig. \ref{FT-attack} shows that even after 50 epochs of fine-tuning, the watermark embedded in the normalization layer by FedZKP can always maintain a 100\% detection rate, which can fully guarantee the security of the verification process. 

\noindent\textbf{Robustness against Pruning Attack:}
The pruning attack \cite{see2016compression} removes redundant parameters from
the trained model. We evaluate the main task performance and watermark detection rate under pruning attack with varying pruning rates.
As shown in Fig. \ref{Purning-atttack}, our FedZKP (green line) can still remain intact under varying purging proportions of network parameters, and the detection rate is maintained at 100\%, which is above the security bound (line with red color) provided in Sect. \ref{Sect:Security Boundary}. 

\begin{figure}[htbp]
    
    \centering
    \subfigure[AlexNet with 5 Clients]{
        \begin{minipage}[t]{0.5\linewidth}
        \centering
        \includegraphics[width=1.7in]{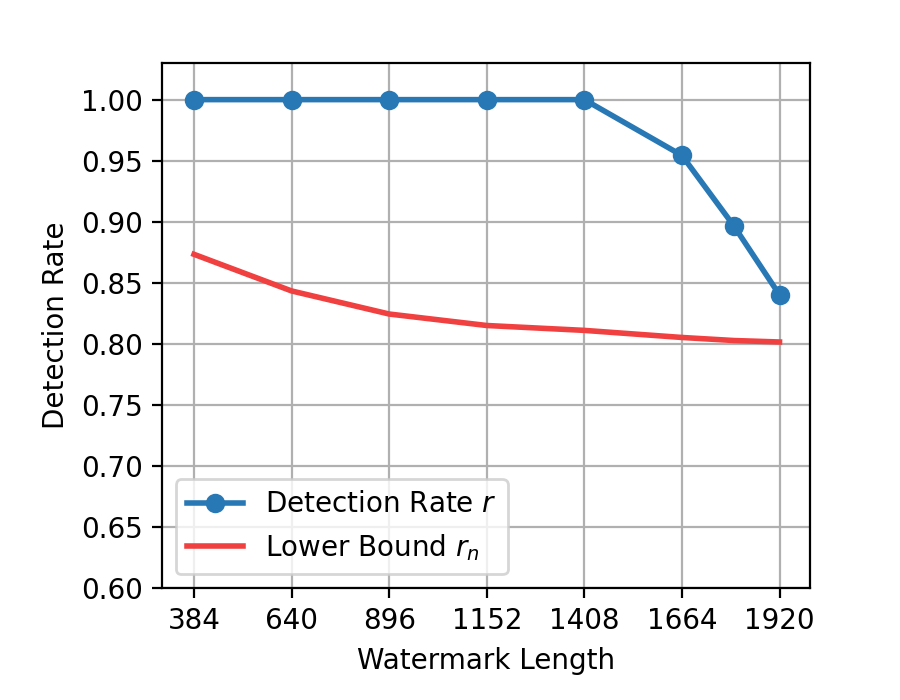}
        \end{minipage}%
    }%
    \subfigure[AlexNet with 10 Clients]{
        \begin{minipage}[t]{0.5\linewidth}
        \centering
        \includegraphics[width=1.7in]{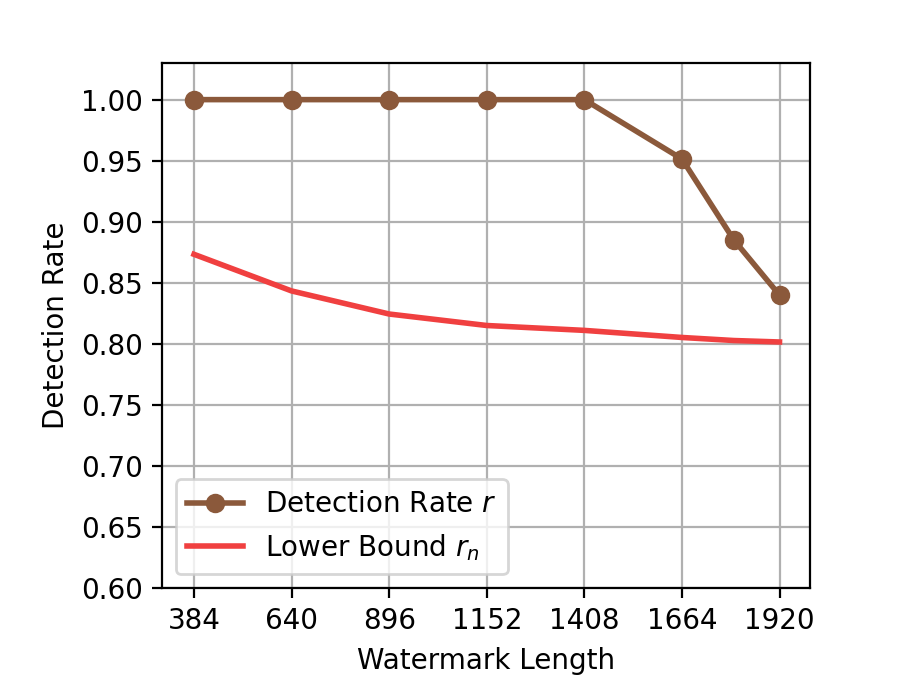}
        \end{minipage}%
    }%
    
    \subfigure[AlexNet with 20 Clients]{
        \begin{minipage}[t]{0.5\linewidth}
        \centering
        \includegraphics[width=1.7in]{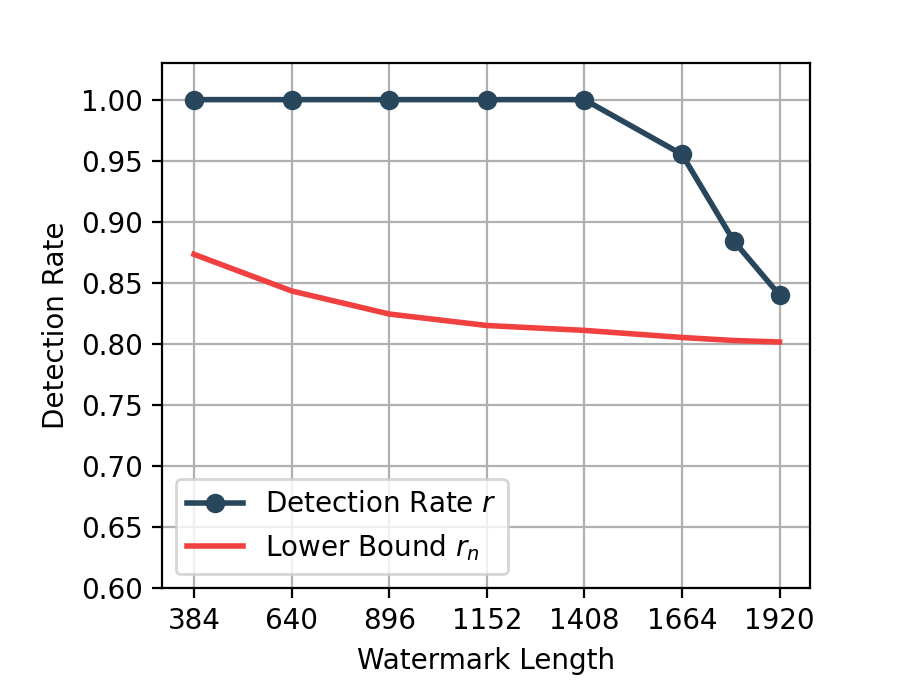}
        \end{minipage}%
    }%
    \subfigure[ResNet with 5 Clients]{
        \begin{minipage}[t]{0.5\linewidth}
        \centering
        \includegraphics[width=1.7in]{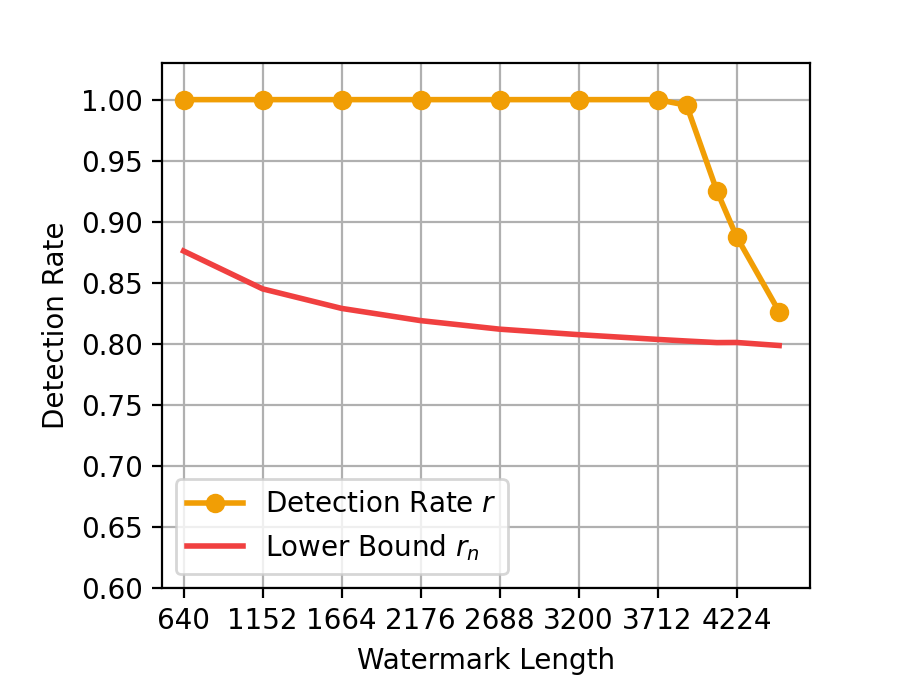}
        \end{minipage}%
    }%

    \subfigure[ResNet with 10 Clients]{
        \begin{minipage}[t]{0.5\linewidth}
        \centering
        \includegraphics[width=1.7in]{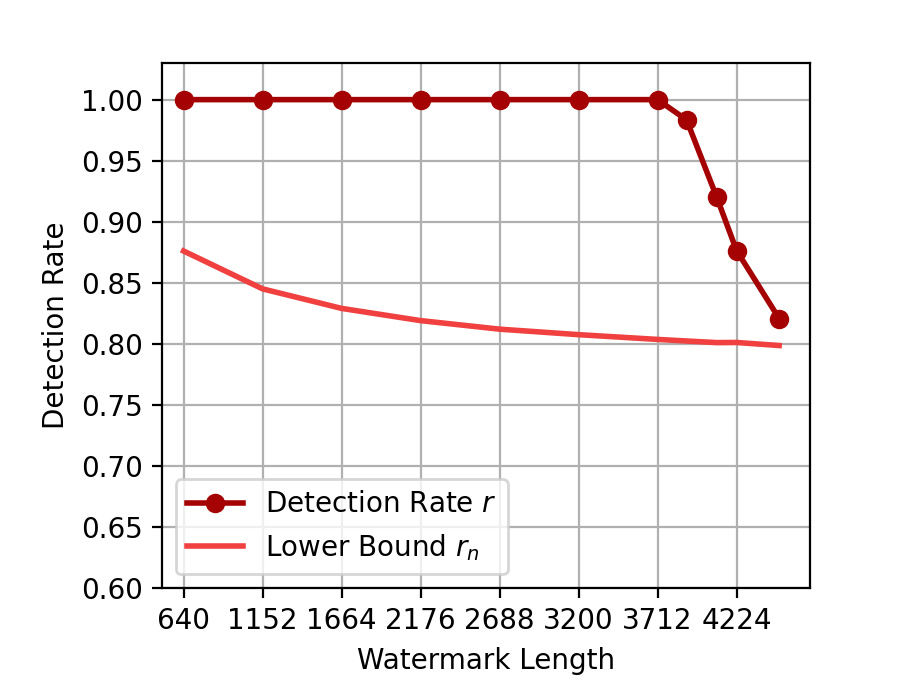}
        \end{minipage}%
    }%
    \subfigure[ResNet with 20 Clients]{
        \begin{minipage}[t]{0.5\linewidth}
        \centering
        \includegraphics[width=1.7in]{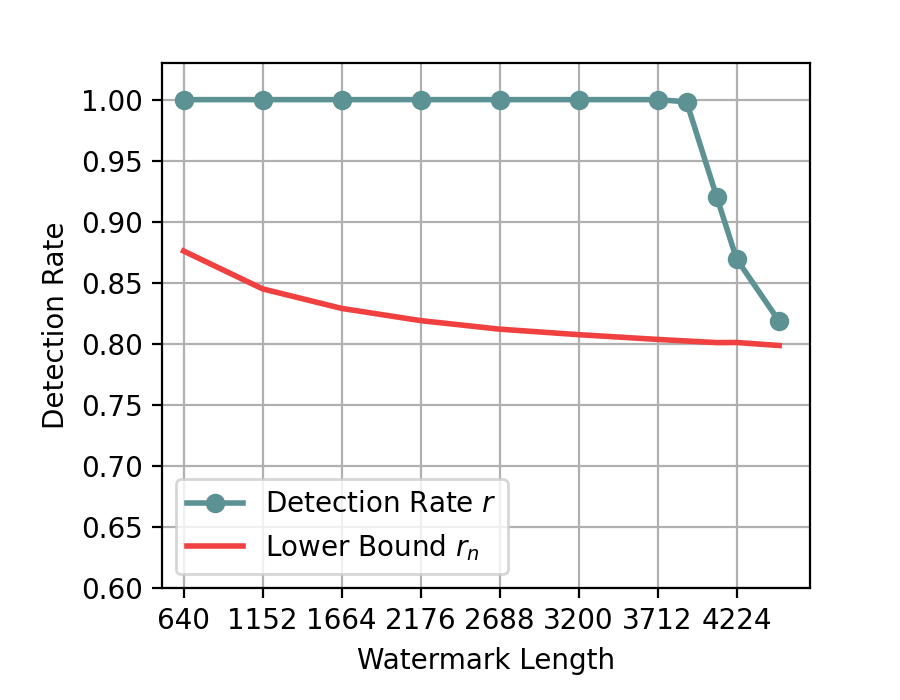}
        \end{minipage}%
    }%
    \centering
    \caption{Figure (a)-(f) shows the watermark detection rate of AlexNet and ResNet18 in various watermark lengths with 5, 10, and 20 clients.}
    \label{Embedded_effect}
\end{figure}

\begin{figure*}
    \centering
    \subfigure[Single Round Time Cost]{
        \includegraphics[scale=0.33]{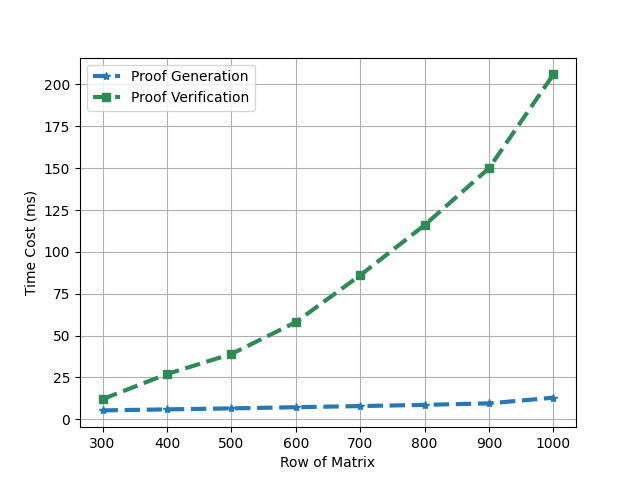} 
    }
    \subfigure[Multi Round Generating Time Cost]{
        \includegraphics[scale=0.33]{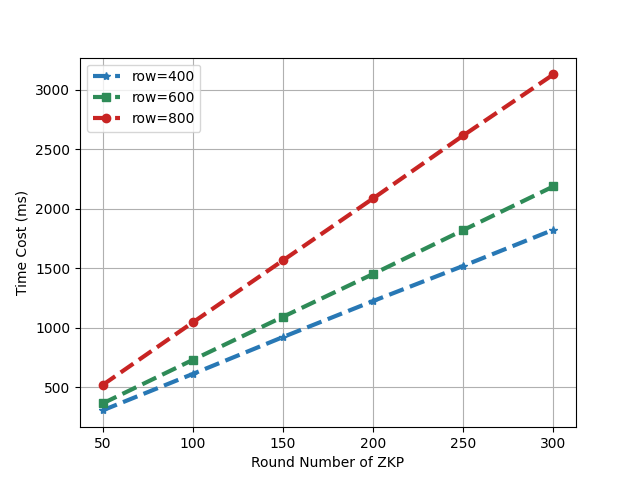} 
    }
    \subfigure[Multi Round Verifying Time Cost]{
        \includegraphics[scale=0.33]{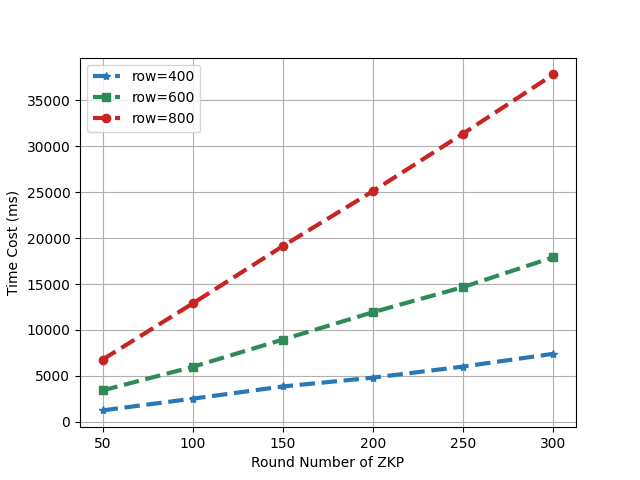}
    }
    \DeclareGraphicsExtensions.
    \caption{Figure (a) represents the relationship between the number of matrix rows and the time cost under a single round of xLPN zero-knowledge proof. Figures (b) and (c) represent the time cost of generating and verifying proof under multiple rounds of ZKP, where rows of matrix $\mathbf{A}_{zkp}$ are 400, 600, and 800, respectively.}
    \label{Time Cost in xLPN ZKP}
\end{figure*}

\begin{figure}
    \centering
    \subfigure[Generating Time Cost]{
        \includegraphics[scale=0.24]{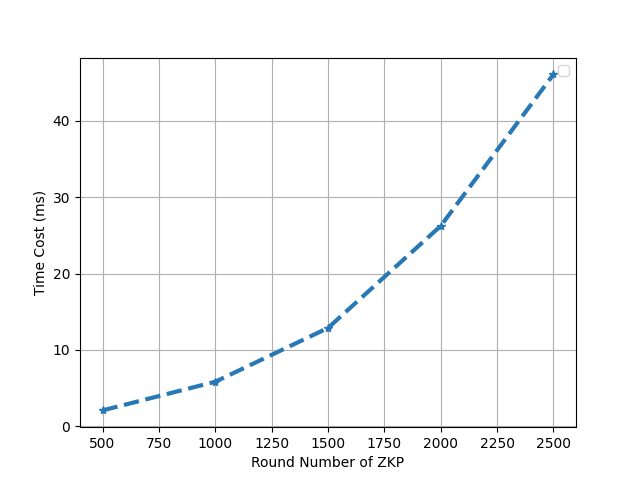} 
    }
    \subfigure[Verifying Time Cost]{
        \includegraphics[scale=0.24]{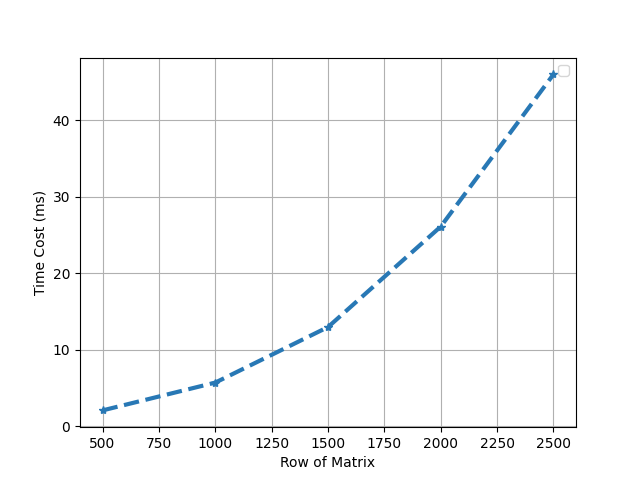} 
    }
    \DeclareGraphicsExtensions.
    \caption{Figures (a) and (b) represent the time cost of generating and verifying commitments.}
    \label{Time Cost in xLPN Commitment}
\end{figure}

\noindent\textbf{Robustness against Targeted Destruction Attack}: The attacker aims to remove the embedding watermark by knowing the way to embed watermarks (e.g., embedding position $\mathcal P$, the hash watermark $h$, and embedding matrix $\mathbf E$). Concretely, the attackers modify the batch normalization parameters towards removing the embedding watermark as the following:

For watermarks  $\mathbf W_{\gamma}=(\gamma^{(1)},\cdots,\gamma^{(\omega)})$ with mean $\mu = \frac{1}{\omega} \sum\limits^{\omega} \limits _{i=1} \gamma^{(i)}$ and variance $\sigma ^2=\frac{1}{\omega} \sum\limits^{\omega} \limits _{i=1} {(\gamma^{(i)}-\mu)}^2$ , we add the Gaussian noise $\upsilon$ to $\mathbf W_\gamma$ as:
\begin{equation}
   \mathbf {W}_{err}=\mathbf W_\gamma+ \Upsilon    
\end{equation}
in which , ${\Upsilon}=(\upsilon ^{(1)},\cdots, \upsilon ^{(\omega)})$ , $\upsilon ^{(i)} \sim \mathcal N(\mu,\varphi \sigma^2)$, $\varphi  \in (0,1)$ is the noise intensity and $\mathcal N$ represents a Gaussian distribution. 

Fig. \ref{Target_Res_1024}-\ref{Target_Alex_1024} shows the robustness of ResNet18 resisting targeted destruction attack with the different watermark length of 1024 bits 2048 bits and 3076 bits, which demonstrate the model performance is influenced although the watermark detection rate decreases with the noise intensity. For instance, On ReNet18 with 2048bits, when the attacker destroys the watermark so that its detection rate is lower than the lower bound, the accuracy of the model drops from 94.1\% to 78.6\%, a loss of 15.5\%. This loss is bigger (19\%) on ReNet18 with 3076bits. Therefore, if the adversary performs a targeted destruction attack on the model, he destroys the watermark at the cost of a serious drop in model performance. Specifically, with the acceptable model performance drop e.g., 10\%, the detection rate is above the security bound provided in Sect. \ref{Sect:Security Boundary}.

\subsubsection{Robustness in Federated Training}
\label{embedded_effect}
We evaluate whether watermarks can be embedded accurately via watermark detection rate $r$ during federated training with different client numbers and lengths of watermarks. From Fig. \ref{Embedded_effect}, we obtain these three observations:

\begin{itemize}
    \item  First, when the watermark length increases to a threshold (e.g., 1408 bits for AlexNet and 3712 bits for ResNet), the detection rate decreases. This is because the dimension of model parameters is limited so that it doesn't support infinity watermarks embedded in the model, which is analyzed in Theorem 1 \cite{li2022fedipr}. Besides, when the length does not exceed the threshold, the watermark can be completely embedded within 5 epochs, i.e., the detection rate reaches 100\%.
    
    \item Second, the detection rate for FedZKP is stable (i.e.,  the difference between the detection rate with 5, 10, and 20 clients is not more than 0.5\%.) with the different number of clients. The reason is that the FedZKP watermark is globally shared and unique, no matter how the number of clients changes, the length of the watermark can be fixed, so changes in the number of clients have very little effect on watermark embedding.
    
    \item Third, the theoretical value of the lower bound (red line) for the detection rate provided in Sect. \ref{Sect:Security Boundary} is less than the experimental detection rate, i.e., the security is guaranteed when the embedding watermark bit length is less than 1664 and 4500 for AlexNet and ResNet respectively.
\end{itemize}

\subsection{Cost in xLPN ZKP  and Commitment}
\subsubsection{Time Cost}
Fig. \ref{Time Cost in xLPN ZKP} illustrates the time cost in single-round ZKP, multi-round proof generation, and multi-round proof verification based on the xLPN problem. In this experiment, the number of columns of the matrix $\mathbf{A}_{zkp}$ is the number of rows minus 100. The time cost of generating the proof grows as the square of the number of rows of the matrix $\mathbf{A}_{zkp}$. Because the verification proof requires solving the system of equations, the time cost of verification grows cubically of the number of rows of the matrix $\mathbf{A}_{zkp}$. But, even if the row number reaches 700, it only takes 8 ms to generate the proof and 86 ms to verify it. With 600 row number of $\mathbf{A}_{zkp}$, 300 rounds of xLPN ZKP interactions, the attacker's knowledge error is $(\frac{2}{3})^{300}$, it takes only 2186 ms to generate the proof and 17919 ms to verify it. The time cost is acceptable for ownership verification.

The time cost of the xLPN commitment scheme \cite{jain2012commitments} used in our scheme are shown in Fig. \ref{Time Cost in xLPN Commitment}. The time overhead for generating and verifying commitments is approximately the same, and it takes only a few milliseconds to generate and verify commitments, which is a good indication of the efficiency of the commitment scheme.

\subsubsection{Memory and Communication Cost}
Suppose there are $K$ clients in the FL system. $K\cdot m\cdot l$ is the size of $\mathbf{A}_{agg}$, $N\cdot l$ is the size of $\mathbf{y}_{agg}$, $l$ is the size of $\mathbf{e},\mathbf{t}_0,\mathbf{t}_1,\mathbf{t}_2$, $m$ is the size of $\mathbf{s}$, the round of verification is $d$. $l_{com}$ is the length of commitment.

The client needs to store the aggregated public input $\mathbf{A}_{agg}, \mathbf{y}_{agg}$, private input $\mathbf{e}, \mathbf{s}$ and the information $\pi,\mathbf{t}_0,\mathbf{t}_1,\mathbf{t}_2$ of commitments in each round of verification. There are $l$ position mapping in the permutation $\pi$, and $l$ integers need to be saved, so $32l$ bits are needed to store the $\pi$. The memory cost is $Nml+m+l+Nl+d(32l+3l)=(Nm+35d+N+1)l+m$ bits. If $m=800, l=700, N=10, d=300$, the memory cost of verification is 1.54MB, which is negligible compared to restoring the  model. 

The communication cost is divided into three parts: send $\mathbf{A}_{agg}, \mathbf{y}_{agg}$, send commitments, and open commitments. The size of $\mathbf{A}_{agg}, \mathbf{y}_{agg}$ is $Nml+Nl$. The size of commitments is $3dl_{com}$. The size of the opening commitment is $(32+3l)d\cdot\frac 2 3$, we only open two of three commitments in a single round, so we should multiply $\frac 2 3$. For $m=800, l=700, l_{com}=800, N=10, d=300$, the communication cost of verification is 824KB, which is also negligible compared to sending the federated model during verification. 
\section{Conclusion}
\label{Sect:Conclusion}
We propose the first zero-knowledge FMOV scheme based on xLPN ZKP called FedZKP. FedZKP doesn't require a trusted verifier to protect the confidentiality of the credentials in FMOV, which reduces the risk of credential leakage. We construct a security model for FMOV, and the formal security proof shows that FedZKP can resist any attack that can be captured by the security model. We give the solution to the technical challenge caused by errors in watermarking caused by model training and removal attack. Security analysis and experimental results show that it is not feasible for the attacker to affect the correctness of ownership verification without excessive degradation of model accuracy. We hope to use cryptography in the future to further accelerate the application of model ownership protection.

{\footnotesize
\bibliographystyle{unsrt}
\bibliography{reference}}

\begin{thebibliography}{10}

\bibitem{mcmahan2016federated}
H~Brendan McMahan, Eider Moore, Daniel Ramage, and Blaise~Ag{\"u}era y~Arcas.
\newblock Federated learning of deep networks using model averaging.
\newblock {\em arXiv preprint arXiv:1602.05629}, 2016.

\bibitem{wang2021voxpopuli}
Changhan Wang, Morgane Riviere, Ann Lee, Anne Wu, Chaitanya Talnikar, Daniel
  Haziza, Mary Williamson, Juan Pino, and Emmanuel Dupoux.
\newblock Voxpopuli: A large-scale multilingual speech corpus for
  representation learning, semi-supervised learning and interpretation.
\newblock In {\em Proceedings of Annual Meeting of the Association for
  Computational Linguistics (ACL)}, pages 993--1003, 2021.

\bibitem{deng2009imagenet}
Jia Deng, Wei Dong, Richard Socher, Li-Jia Li, Kai Li, and Li~Fei-Fei.
\newblock Imagenet: A large-scale hierarchical image database.
\newblock In {\em Proceedings of the IEEE conference on Computer Vision and
  Pattern Recognition (CVPR)}, pages 248--255, 2009.

\bibitem{chelba2013one}
Ciprian Chelba, Tomas Mikolov, Mike Schuster, Qi~Ge, Thorsten Brants, Phillipp
  Koehn, and Tony Robinson.
\newblock One billion word benchmark for measuring progress in statistical
  language modeling.
\newblock {\em arXiv preprint arXiv:1312.3005}, 2013.

\bibitem{zhu2015aligning}
Yukun Zhu, Ryan Kiros, Rich Zemel, Ruslan Salakhutdinov, Raquel Urtasun,
  Antonio Torralba, and Sanja Fidler.
\newblock Aligning books and movies: Towards story-like visual explanations by
  watching movies and reading books.
\newblock In {\em Proceedings of the IEEE conference on Computer Vision and
  Pattern Recognition (CVPR)}, pages 19--27, 2015.

\bibitem{li2022fedipr}
Bowen Li, Lixin Fan, Hanlin Gu, Jie Li, and Qiang Yang.
\newblock Fedipr: Ownership verification for federated deep neural network
  models.
\newblock {\em IEEE Transactions on Pattern Analysis and Machine Intelligence
  (PAMI)}, 2022.

\bibitem{tekgul2021waffle}
Buse~GA Tekgul, Yuxi Xia, Samuel Marchal, and N~Asokan.
\newblock Waffle: Watermarking in federated learning.
\newblock In {\em Proceedings of International Symposium on Reliable
  Distributed Systems (SRDS)}, pages 310--320. IEEE, 2021.

\bibitem{jain2012commitments}
Abhishek Jain, Stephan Krenn, Krzysztof Pietrzak, and Aris Tentes.
\newblock Commitments and efficient zero-knowledge proofs from learning parity
  with noise.
\newblock In {\em Proceedings of International Conference on the Theory and
  Application of Cryptology and Information Security (Asiacrypt)}, pages
  663--680, 2012.

\bibitem{goldwasser2019knowledge}
Shafi Goldwasser, Silvio Micali, and Chales Rackoff.
\newblock The knowledge complexity of interactive proof-systems.
\newblock In {\em Providing Sound Foundations for Cryptography: On the Work of
  Shafi Goldwasser and Silvio Micali}, pages 203--225. 2019.

\bibitem{mcmahan2017communication}
Brendan McMahan, Eider Moore, Daniel Ramage, Seth Hampson, and Blaise~Aguera
  y~Arcas.
\newblock Communication-efficient learning of deep networks from decentralized
  data.
\newblock In {\em Proceedings of International Conference on Artificial
  Intelligence and Statistics (AISTATS)}, pages 1273--1282, 2017.

\bibitem{konevcny2016federated}
Jakub Kone{\v{c}}n{\`y}, H~Brendan McMahan, Daniel Ramage, and Peter
  Richt{\'a}rik.
\newblock Federated optimization: Distributed machine learning for on-device
  intelligence.
\newblock {\em arXiv preprint arXiv:1610.02527}, 2016.

\bibitem{yang2019federated}
Qiang Yang, Yang Liu, Tianjian Chen, and Yongxin Tong.
\newblock Federated machine learning: Concept and applications.
\newblock {\em ACM Transactions on Intelligent Systems and Technology (TIST)},
  10(2):1--19, 2019.

\bibitem{yang2023federated}
Qiang Yang, Anbu Huang, Lixin Fan, Chee~Seng Chan, Jian~Han Lim, Kam~Woh Ng,
  Ding~Sheng Ong, and Bowen Li.
\newblock Federated learning with privacy-preserving and model
  ip-right-protection.
\newblock {\em Machine Intelligence Research}, 20(1):19--37, 2023.

\bibitem{10.1145/3433210.3437526}
Xiaoyu Cao, Jinyuan Jia, and Neil~Zhenqiang Gong.
\newblock Ipguard: Protecting intellectual property of deep neural networks via
  fingerprinting the classification boundary.
\newblock In {\em Proceedings of ACM Asia Conference on Computer and
  Communications Security (AsiaCCS)}, page 14–25, 2021.

\bibitem{li2021modeldiff}
Yuanchun Li, Ziqi Zhang, Bingyan Liu, Ziyue Yang, and Yunxin Liu.
\newblock Modeldiff: testing-based dnn similarity comparison for model reuse
  detection.
\newblock In {\em Proceedings of the International Symposium on Software
  Testing and Analysis (ISSTA)}, pages 139--151, 2021.

\bibitem{zhao2020afa}
Jingjing Zhao, Qingyue Hu, Gaoyang Liu, Xiaoqiang Ma, Fei Chen, and
  Mohammad~Mehedi Hassan.
\newblock Afa: Adversarial fingerprinting authentication for deep neural
  networks.
\newblock {\em Computer Communications}, 150:488--497, 2020.

\bibitem{jia2021proof}
Hengrui Jia, Mohammad Yaghini, Christopher~A Choquette-Choo, Natalie Dullerud,
  Anvith Thudi, Varun Chandrasekaran, and Nicolas Papernot.
\newblock Proof-of-learning: Definitions and practice.
\newblock In {\em Proceedings of IEEE Symposium on Security and Privacy
  (S\&P)}, pages 1039--1056, 2021.

\bibitem{uchida2017embedding}
Yusuke Uchida, Yuki Nagai, Shigeyuki Sakazawa, and Shin'ichi Satoh.
\newblock Embedding watermarks into deep neural networks.
\newblock In {\em Proceedings of ACM SIGMM International Conference on
  Multimedia Retrieva (ICMR)}, pages 269--277, 2017.

\bibitem{fan2019rethinking}
Lixin Fan, Kam~Woh Ng, and Chee~Seng Chan.
\newblock Rethinking deep neural network ownership verification: Embedding
  passports to defeat ambiguity attacks.
\newblock {\em Advances in neural information processing systems}, 32, 2019.

\bibitem{fan2021deepip}
Lixin Fan, Kam~Woh Ng, Chee~Seng Chan, and Qiang Yang.
\newblock Deepip: Deep neural network intellectual property protection with
  passports.
\newblock {\em IEEE Transactions on Pattern Analysis and Machine Intelligence
  (PAMI)}, (01):1--1, 2021.

\bibitem{chen2018deepmarks}
Huili Chen, Bita~Darvish Rohani, and Farinaz Koushanfar.
\newblock Deepmarks: A digital fingerprinting framework for deep neural
  networks.
\newblock {\em arXiv preprint arXiv:1804.03648}, 2018.

\bibitem{zhang2020passport}
Jie Zhang, Dongdong Chen, Jing Liao, Weiming Zhang, Gang Hua, and Nenghai Yu.
\newblock Passport-aware normalization for deep model protection.
\newblock {\em Advances in Neural Information Processing Systems},
  33:22619--22628, 2020.

\bibitem{adi2018turning}
Yossi Adi, Carsten Baum, Moustapha Cisse, Benny Pinkas, and Joseph Keshet.
\newblock Turning your weakness into a strength: Watermarking deep neural
  networks by backdooring.
\newblock In {\em Proceedings of USENIX Security Symposium (USENIX Security)},
  pages 1615--1631, 2018.

\bibitem{zhang2018protecting}
Jialong Zhang, Zhongshu Gu, Jiyong Jang, Hui Wu, Marc~Ph Stoecklin, Heqing
  Huang, and Ian Molloy.
\newblock Protecting intellectual property of deep neural networks with
  watermarking.
\newblock In {\em Proceedings of ACM Asia Conference on Computer and
  Communications Security (AsiaCCS)}, pages 159--172, 2018.

\bibitem{lukas2019deep}
Nils Lukas, Yuxuan Zhang, and Florian Kerschbaum.
\newblock Deep neural network fingerprinting by conferrable adversarial
  examples.
\newblock {\em arXiv preprint arXiv:1912.00888}, 2019.

\bibitem{liu2021secure}
Xiyao Liu, Shuo Shao, Yue Yang, Kangming Wu, Wenyuan Yang, and Hui Fang.
\newblock Secure federated learning model verification: A client-side backdoor
  triggered watermarking scheme.
\newblock In {\em Proceedings of IEEE International Conference on Systems, Man,
  and Cybernetics (SMC)}, pages 2414--2419. IEEE, 2021.

\bibitem{veron1997improved}
Pascal V{\'e}ron.
\newblock Improved identification schemes based on error-correcting codes.
\newblock {\em Applicable Algebra in Engineering, Communication and Computing},
  8(1):57--69, 1997.

\bibitem{ben2013snarks}
Eli Ben-Sasson, Alessandro Chiesa, Daniel Genkin, Eran Tromer, and Madars
  Virza.
\newblock Snarks for c: Verifying program executions succinctly and in zero
  knowledge.
\newblock In {\em Annual cryptology conference}, pages 90--108, 2013.

\bibitem{groth2016size}
Jens Groth.
\newblock On the size of pairing-based non-interactive arguments.
\newblock In {\em Proceedings of International Conference on the Theory and
  Applications of Cryptographic Techniques (EUROCRYPT)}, pages 305--326, 2016.

\bibitem{ames2017ligero}
Scott Ames, Carmit Hazay, Yuval Ishai, and Muthuramakrishnan
  Venkitasubramaniam.
\newblock Ligero: Lightweight sublinear arguments without a trusted setup.
\newblock In {\em Proceedings of ACM Conference on Computer and Communications
  Security (CCS)}, pages 2087--2104, 2017.

\bibitem{bunz2018bulletproofs}
Benedikt B{\"u}nz, Jonathan Bootle, Dan Boneh, Andrew Poelstra, Pieter Wuille,
  and Greg Maxwell.
\newblock Bulletproofs: Short proofs for confidential transactions and more.
\newblock In {\em Proceedings of IEEE Symposium on Security and Privacy
  (S\&P)}, pages 315--334, 2018.

\bibitem{lee2020vcnn}
Seunghwa Lee, Hankyung Ko, Jihye Kim, and Hyunok Oh.
\newblock vcnn: Verifiable convolutional neural network based on zk-snarks.
\newblock {\em Cryptology ePrint Archive}, 2020.

\bibitem{weng2021mystique}
Chenkai Weng, Kang Yang, Xiang Xie, Jonathan Katz, and Xiao Wang.
\newblock Mystique: Efficient conversions for $\{$Zero-Knowledge$\}$ proofs
  with applications to machine learning.
\newblock In {\em Proceedings of USENIX Security Symposium (USENIX Security)},
  pages 501--518, 2021.

\bibitem{damgaard2002sigma}
Ivan Damg{\aa}rd.
\newblock On $\sigma$-protocols.
\newblock {\em Lecture Notes, University of Aarhus, Department for Computer
  Science}, page~84, 2002.

\bibitem{menezes2001handbook}
Alfred~J Menezes, Paul~C Van~Oorschot, and Scott~A Vanstone.
\newblock {\em Handbook of applied cryptography}.
\newblock chapter Hash Functions and Data Integrity, 2001.

\bibitem{rosasco2004loss}
Lorenzo Rosasco, Ernesto De~Vito, Andrea Caponnetto, Michele Piana, and
  Alessandro Verri.
\newblock Are loss functions all the same?
\newblock {\em Neural computation}, 16(5):1063--1076, 2004.

\bibitem{bellare1993random}
Mihir Bellare and Phillip Rogaway.
\newblock Random oracles are practical: A paradigm for designing efficient
  protocols.
\newblock In {\em Proceedings of ACM Conference on Computer and Communications
  Security (CCS)}, pages 62--73, 1993.

\bibitem{krizhevsky2017imagenet}
Alex Krizhevsky, Ilya Sutskever, and Geoffrey~E Hinton.
\newblock Imagenet classification with deep convolutional neural networks.
\newblock {\em Communications of the ACM}, 60(6):84--90, 2017.

\bibitem{he2016deep}
Kaiming He, Xiangyu Zhang, Shaoqing Ren, and Jian Sun.
\newblock Deep residual learning for image recognition.
\newblock In {\em Proceedings of the IEEE conference on Computer Vision and
  Pattern Recognition (CVPR)}, pages 770--778, 2016.

\bibitem{krizhevsky2009learning}
Alex Krizhevsky, Geoffrey Hinton, et~al.
\newblock Learning multiple layers of features from tiny images.
\newblock 2009.

\bibitem{dworkin2015sha}
Morris~J Dworkin.
\newblock Sha-3 standard: Permutation-based hash and extendable-output
  functions.
\newblock 2015.

\bibitem{levieil2006improved}
{\'E}ric Levieil and Pierre-Alain Fouque.
\newblock An improved lpn algorithm.
\newblock In {\em Proceedings of Security and Cryptography for Networks}, pages
  348--359, 2006.

\bibitem{pietrzak2012cryptography}
Krzysztof Pietrzak.
\newblock Cryptography from learning parity with noise.
\newblock In {\em SOFSEM}, volume~12, pages 99--114, 2012.

\bibitem{simonyan2014very}
Karen Simonyan and Andrew Zisserman.
\newblock Very deep convolutional networks for large-scale image recognition.
\newblock {\em arXiv preprint arXiv:1409.1556}, 2014.

\bibitem{see2016compression}
Abigail See, Minh-Thang Luong, and Christopher~D Manning.
\newblock Compression of neural machine translation models via pruning.
\newblock {\em arXiv preprint arXiv:1606.09274}, 2016.

\end{thebibliography}
\FloatBarrier
\appendix
\section{Experimental Setting Details}
\label{Sect:AppendixExp}
The hyperparameter settings for federated AlexNet and ResNet18 can be seen in Table \ref{Table:Traning Parameters Alex} and Table \ref{Table:Traning Parameters Res}.
The specific architectures of AlexNet and ResNet18 are shown in Table \ref{Table:Alex} and Table \ref{Table:ResNet18}.

\begin{table}[]
\centering
\caption{Training parameters for Federated AlexNet}
\begin{tabular}{c|c}
\hline
Hyper-parameter     & AlexNet                                                 \\ \hline
Optimization method & SGD                                                         \\
Learning rate       & 0.01                                                       \\
Batch size          & 16                                                              \\
Global Epochs       & 200                                                    \\
Local Epochs        & 2                                                                   \\
Learning rate decay & 0.99 at each global Epoch  \\
Regularization Term & BCE loss, Hinge-like loss     \\ \hline

\end{tabular}
\label{Table:Traning Parameters Alex}
\end{table}

\begin{table}[]
\centering
\caption{Training parameters for Federated ResNet18}
\begin{tabular}{c|c}
\hline
Hyper-parameter                  & ResNet-18                                   \\ \hline
Optimization method            & SGD                                              \\
Learning rate                      & 0.01                                             \\
Batch size                           & 16                                                 \\
Global Epochs                        & 200                                                \\
Local Epochs                         & 2                                                 \\
Learning rate decay  & 0.100 at each global Epoch  \\
Regularization Term  & BCE loss, Hinge-like loss     \\ \hline

\end{tabular}
\label{Table:Traning Parameters Res}
\end{table}

\begin{table}[]
\caption {AlexNet Architecture used in FedZKP (Embed feature-based
watermarks across Conv3, Conv4 and Conv5)}
\begin{tabular}{c|c|c|c}
\hline
Layer name & Output size  & Weight shape                     & Padding \\ \hline
Conv1      & 32 $\times$ 32 & 64 $\times$ 3$\times$ 5$\times$ 5      & 2       \\
MaxPool2d  & 16 $\times$ 16 & 2 $\times$ 2                       &         \\
Conv2      & 16 $\times$ 16 & 192 $\times$ 64 $\times$ 5 $\times$ 5  & 2       \\
Maxpool2d  & 8 $\times$ 8   & 2 $\times$ 2                       &         \\
Conv3      & 8 $\times$ 8   & 384 $\times$ 192 $\times$ 3 $\times$ 3 & 1       \\
Conv4      & 8 $\times$ 8   & 256$\times$ 384 $\times$ 3 $\times$ 3  & 1       \\
Conv5      & 8 $\times$ 8   & 256 $\times$ 256 $\times$ 3 $\times$ 3 &         \\
MaxPool2d  & 4 $\times$ 4   & 2 $\times$ 2                       &         \\
Linear     & 10           & 10 $\times$ 4096                   &         \\ \hline
\end{tabular}
\label{Table:Alex}
\end{table}

\begin{table}[]
\caption {ResNet18 Architecture used in FedZKP (Embed
watermarks across Res5 Block)}
\begin{tabular}{c|c|c|c}
\hline
Layer name & Output size  & Weight shape                     & Padding \\ \hline
Conv1      & 32 $\times$32 & 64 $\times$3$\times$5$\times$5      & 1       \\ \hline
Res2       & 32 $\times$32 & $\left [ \begin{matrix}
64 \times 64 \times 3 \times 3 \\
64\times64\times3\times3 \\
\end{matrix} \right ]$ $\times$2                    & 1       \\ \hline
Res3       & 16 $\times$16 & $\left [ \begin{matrix}
128\times128\times3\times3 \\
128\times128\times3\times3 \\
\end{matrix} \right ] $ $\times$2  & 1       \\ \hline
Res4       & 8 $\times$8   & $\left [ \begin{matrix}
256\times256\times3\times3 \\
256\times256\times3\times3 \\
\end{matrix} \right ]$ $\times$2                   & 1       \\ \hline
Res5       & 4 $\times$4   & $\left [ \begin{matrix}
512\times512\times3\times3 \\
512\times512\times3\times3 \\
\end{matrix} \right ]$ $\times$2 & 1       \\ \hline
Linear     & 100          & 100 $\times$512                   &         \\ \hline
\end{tabular}
\label{Table:ResNet18}
\end{table}
\FloatBarrier
\end{document}